\documentclass[useAMS,usenatbib,usegraphicx]{mn2e}

\usepackage{url}
\usepackage[varg]{txfonts}

\graphicspath{{Figures/}}

\DeclareRobustCommand{\ion}[2]{%
  \textup{#1\,{\mdseries\textsc{#2}}}%
}

\renewcommand\vec[1]{\bmath{#1}}

\title[Two-dimensional kinematics of SLACS lenses --
IV.]{Two-dimensional kinematics of SLACS lenses -- IV. The complete
  VLT--VIMOS data set\thanks{Based on observations made with ESO
    Telescopes at the La Silla or Paranal Observatories under
    programme ID 075.B-0226 and 177.B-0682 and on observations made
    with the NASA/ESA Hubble Space Telescope, obtained at the Space
    Telescope Science Institute, which is operated by the Association
    of Universities for Research in Astronomy, Inc., under NASA
    contract NAS 5-26555.}}

\author[O. Czoske et al.]{%
  Oliver
  Czoske$^{1,2}$\thanks{E-mail:~oliver.czoske@univie.ac.at}, Matteo
  Barnab\`e$^{3,2}$, L\'eon V. E. Koopmans$^{2}$,
  Tommaso Treu$^{4}$ \newauthor
  and Adam S. Bolton$^{5}$\\
  $^{1}$~Institut f\"ur Astronomie der Universit\"at Wien,
  T\"urkenschanzstra\ss e 17, 1180 Wien, Austria\\
  $^{2}$~Kapteyn Astronomical Institute, P.\,O. Box 800, 9700\,AV
  Groningen, The Netherlands\\
  $^{3}$~Kavli Institute for Particle
  Astrophysics and Cosmology, Stanford University, 452 Lomita Mall,
  Stanford, CA 94035-4085, USA\\
  $^{4}$~Department of Physics, University of California, Santa
  Barbara, CA 93101, USA\\
  $^{5}$~Department of Physics and Astronomy, University of Utah, 115
  South 1400 East, Salt Lake City, UT 84112, USA}

\begin{document}


\pagerange{\pageref{firstpage}--\pageref{lastpage}} \pubyear{2011}

\maketitle

\label{firstpage}

\begin{abstract}
  This paper presents the full VLT/VIMOS-IFU data set and related data
  products from an ESO Large Programme with the observational goal of
  obtaining two-dimensional kinematic data of early-type lens
  galaxies, out to one effective radius. The sample consists of
  17 early-type galaxies (ETG) selected from the SLACS gravitational-lens
  survey. The galaxies cover the redshift range from 0.08 to 0.35 and
  have stellar velocity dispersions between 200 and 350 km/s. This
  programme is complemented by a similar observational programme on Keck,
  using long-slit spectroscopy.  In combination with multi-band
  imaging data, the kinematic data provide stringent constraints on
  the inner mass profiles of ETGs beyond the local universe. Our Large
  Programme thus extends studies of nearby early-type galaxies (e.g.\
  SAURON/ATLAS3D) by an order of magnitude in distance and toward
  higher masses.
  We provide an overview of our observational strategy, the data
  products (luminosity-weighted spectra and Hubble Space Telescope
  images) and derived products (i.e.\ two-dimensional fields of
  velocity dispersions and streaming motions) that have been used in a
  number of published and forthcoming lensing, kinematic and
  stellar-population studies.
  These studies also pave the way for future studies of early-type
  galaxies at $z\approx 1$ with the upcoming extremely large
  telescopes.
\end{abstract}


\section{Introduction}
\label{sec:Introduction}

The formation and evolution of early-type galaxies (ETGs) is a widely
studied topic in present-day astrophysics, in particular due to a
number of tight correlations between their observables, such as
structural parameters, kinematics, colours and central supermassive
black-hole masses \citep[e.g.][]{Illingworth1977,
  Sandage-Visvanathan1978, Djorgovski-Davis1987, Dressler1987,
  Ferrarese-Merritt2000, Gebhardt2000}. See \citet{Renzini2006} for a
recent review. These correlations seem hard to explain at first sight
in the hierarchical CDM formation paradigm \citep[e.g.][]{Binney1978,
  Blumenthal1984}, where ETGs are seen as the end products of the
stochastical merging of smaller building blocks \citep[e.g.\ spiral
galaxies;][]{Toomre-Toomre1972, Barnes1992}.

Both observational and theoretical studies have progressed rapidly
over the last decade, largely thanks to major observational programmes
conducted on the ground and from space \citep[e.g.\ 2dF, SDSS, COSMOS,
COMBO17;][]{Colless2001, Bernardi2003short, Scarlata2007short,
  Faber2007short} and due to the large-scale numerical simulations
nowadays affordable by supercomputers \citep[e.g.][]{Hopkins2006}.
The SAURON and ATLAS\textsuperscript{3D} projects provide detailed
analysis on substantial samples of nearby ETGs based on deep
integral-field spectroscopic data and serve as a local benchmark for
ETG studies
\citep[e.g.][]{Cappellari2007,Emsellem2007,Emsellem2011short}.

Despite the tremendous increase in data volume and computational
power, many questions remain very difficult to answer. This is due to
several factors: (i)~the types of available data (usually only
luminosity-weighted kinematics and imaging) and their quality (in
terms of signal-to-noise ratio and spatial/spectral resolution) have
remained limited compared to their volume, (ii)~mass modeling
techniques often suffer from intrinsic degeneracies that are hard to
overcome without very high quality data that can only be obtained for
small samples of nearby ETGs. ETG studies that are now conducted
\textsl{beyond} the local Universe with 8--10\,m-class and space-based
telescopes suffer from many of the same issues that similar studies of
the \textsl{local} Universe, with 4\,m-class telescopes, faced more
than a decade ago.

One of the prevailing and, even today, still not precisely answered
questions is how important dark matter is in the inner
baryon-dominated regions of ETGs \citep[e.g.][]{Saglia1992,
  Saglia1993, Bertola1993, Bertin1994, Carollo1995, Gerhard1998,
  Loewenstein-White1999, Gerhard2001, Keeton2001, Romanowsky2003,
  Mamon-Lokas2005a, deLorenzi2009}.  Open questions include in
particular the following: (i)~How much dark matter is there precisely
inside the inner few effective radii of ETGs and how is it
distributed?  (ii)~How do the stellar and dark matter distributions of
ETGs evolve with cosmic time and are there trends with galaxy mass?
(iii)~Do these observations agree with theoretical predictions?

Because dark matter is not directly observable but can only be
inferred from other observations, these questions are particularly
difficult to answer for more distant ETGs where data quality
progressively deteriorates due to lack of signal-to-noise, even with
the largest ground-based telescopes.  To overcome a number of these
hurdles, several systematic programmes were initiated over the last
decade to combine the constraints of strong gravitational lensing with
those of stellar kinematics \citep[LSD and SLACS;
e.g.][]{Koopmans-Treu2002, Treu-Koopmans2004, Bolton2006, Treu2006,
  Koopmans2006, Koopmans2009, Bolton2008a, Auger2009}.  This
combination has turned out particularly powerful in breaking
degeneracies in ETG mass models, despite the limited quality of the
kinematic data \citep[e.g.][]{Treu-Koopmans2004, Koopmans2006,
  Koopmans2009, Auger2010b}. The reason is that the mass enclosed by
the Einstein radius of a lens galaxy can be accurately determined and
breaks the usual mass-anisotropy degeneracy of ETGs to a large extent
\citep[see e.g.][for an explanation based on a simple power-law
model]{Koopmans2004}.

Whereas the LSD and SLACS programmes used predominantly
luminosity-weighted stellar velocity dispersions and were modeled
using the spherical Jeans equation \citep[e.g.][]{Koopmans-Treu2003,
  Treu-Koopmans2004}, one might argue that (slowly) rotating systems
and/or systems with orbital anisotropy cannot be modeled properly that
way \citep[see for example the discussion in][]{Kochanek2006}.  To
address these valid issues, a new and substantially larger
observational programme of a sub-sample of SLACS lenses, using
observations with the VLT and Keck, was started in 2006
\citep[][]{Czoske2008, Czoske2009short, Barnabe2010} to obtain
two-dimensional kinematic fields (both first and second moments of the
line-of-sight velocity distribution) out to their effective radii,
complemented with a rigorous modeling effort based on two-integral
axisymmetric models \citep[][]{Barnabe-Koopmans2007,
  Barnabe2009a}. The self-consistent combination of
gravitational-lensing and stellar-kinematic data sets allows better
constraints to be set on the mass distribution of ETGs
\citep[][]{Czoske2008, Barnabe2009b, Barnabe2010,Barnabe2011}, but
also to test less rigorous methods that might still have to be used at
even higher redshifts (i.e.\ $z\ga 0.5$) until IFU observations from
the next generation of telescopes (such as ESO's ELT) become
available.

Whereas the above-mentioned results show that these data provide more
precise and accurate results on the \textsl{total} mass profile of
ETGs in the inner few effective radii, several problems still remain
that are critical for our understanding of galaxy formation. For
example, what fraction of the mass inside several effective radii is
due to dark matter and what are the relative mass distributions of the
stellar and dark-matter components? This is much harder to answer with
kinematic and lensing data alone, since there are many combinations of
the stellar and dark-matter distributions that lead to similar lensing
and kinematic data \citep[][]{Treu-Koopmans2004, Barnabe2009b}.  If,
on the other hand, the stellar mass-to-light ratio can be determined
independently, these degeneracies can mostly be lifted.  This assumes,
however, that the broad-band colours provide sufficient information to
break degeneracies in the IMF models, which in general they do
not. Still, current constraints can already set limits on the IMF and
stellar mass-to-light ratios that are becoming competitive
\citep[][]{Grillo2008, Auger2009, Auger2010a, Spiniello2011pre}. To go
beyond these broad-band inferred stellar mass-to-light ratios, one can
also use the same (IFU) spectra to determine line indices from which
more precise ages, metallicities and mass-to-light ratios of the ETG
stellar population can be obtained
\citep[e.g.][]{Trager2000a}. Whereas this information is going to be
used in forthcoming publications, here we concentrate on describing
the data for kinematic purposes.

In this paper, we present the full VLT/VIMOS-IFU data of our sample of
17 SLACS early-type lens galaxies, collected for these purposes. The
sample is presented in Sect.~\ref{sec:Sample}. We describe the
integral-field spectroscopic observations using VIMOS on the VLT in
Sect.~\ref{ssec:obs:VIMOS} and summarize the HST observations as
needed for this project in Sect.~\ref{ssec:obs:HST}. The data
reduction of the IFU data is described in some detail in
Sect.~\ref{sec:data_reduction}, and the kinematic analysis, i.e.~the
derivation of two-dimensional maps of line-of-sight velocity and
velocity dispersion, in Sect.~\ref{sec:kinematic_analysis}.
In Sect.~\ref{sec:other_sources}, we provide a list of additional
objects that are visible in the VIMOS/IFU field of view around the
lens galaxies. Finally, we conclude in
Sect.~\ref{sec:conclusions}. Appendix~\ref{sec:noise_estimation} gives
a recipe for estimating the expected photon noise on the reduced
spectra based on a simple model.

The fully self-consistent combined lensing/dynamics analysis of the
data described in this paper is presented in the companion paper by
\citet{Barnabe2011}.

\section{Sample}
\label{sec:Sample}

The sample described in this paper consists of seventeen early-type
galaxies from the Sloan Digital Sky Survey that have been confirmed as
secure gravitational-lens systems in the SLACS survey.\footnote{In
  fact, one system, J1250B, was downgraded to grade \textit{B}
  (``possible lens'') in \citet{Bolton2008a}.}

The SLACS survey searches for gravitational lens systems in the
spectroscopic data of a subsample of SDSS galaxies comprising the
luminous red galaxy sample \citep{Eisenstein2001short} and passive
members of the MAIN sample \citep{Strauss2002short}, defined as having
rest-frame \ion{H}{$\alpha$} equivalent widths less than
$1.5$\,\AA. The presence of emission lines at a redshift higher than
that of the target galaxy in the SDSS spectra indicates that there is
a background galaxy within the fibre, which may have been
lensed. Candidates were selected for follow-up snapshot imaging with
HST, allowing in most cases a robust confirmation or rejection of the
lensing hypothesis. See \citet{Bolton2006} for details on the
selection procedure. \citet{Treu2006} conducted a number of tests to
verify that the SLACS lens sample is statistically indistinguishable
from the parent samples. Results on the structure of the early-type
lens galaxies can therefore be taken as representative for the
population of (massive) early-type galaxies as a whole.

In order to be observable from Paranal, lens systems close to the
equator ($\delta < 15\degr$) were selected from the (mostly northern)
SLACS catalogue. The final sample of 17 systems chosen for VIMOS/IFU
follow-up is evenly distributed in right ascension.

\begin{figure}
  \centering
  \resizebox{\hsize}{!}{\includegraphics{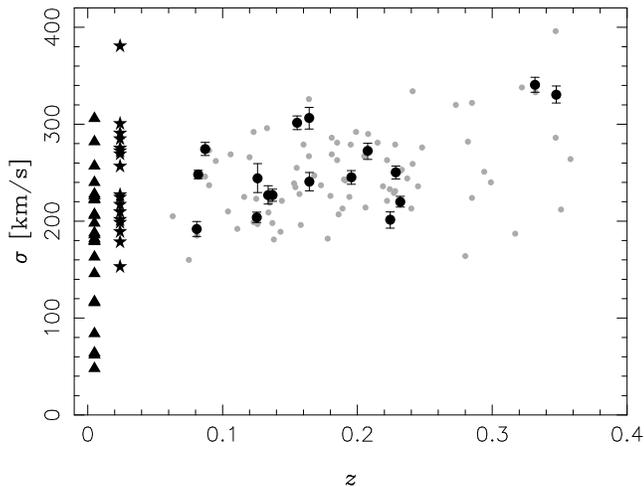}}
  \caption{Distribution of the SLACS/IFU sample in redshift and
    velocity dispersion. The velocity dispersion measurements are
    taken from this paper, the error bars are statistical errors. For
    comparison, the SLACS early-type sample of \citet[grey
    points]{Auger2009}, the SAURON sample at redshift $z\approx 0$
    \citep[triangles:][]{Emsellem2004} and a sample of Coma cluster
    early-type galaxies \citep[stars:][]{Mehlert2000, Corsini2008} are
    shown.}
  \label{fig:Sample_comparison}
\end{figure}

The location of the 84 early-type SLACS galaxies studied by
\citet{Auger2009} in the parameter space of redshift and velocity
dispersion is shown in Fig.~\ref{fig:Sample_comparison}. The seventeen
galaxies from the VIMOS/IFU subsample are marked; they are
representative of the full SLACS early-type galaxy
sample. Fig.~\ref{fig:Sample_comparison} also shows the 48~local
early-type galaxies studied by the SAURON team \citep{Emsellem2004},
as well as 17~early-type galaxies in the Coma cluster studied by
\citet{Thomas2007}. The VIMOS/IFU sample represents a major step out
to cosmological distances compared to the SAURON and Coma samples. In
terms of velocity dispersion, it overlaps with the more massive half
of the SAURON sample and thus lends itself to comparison of the
structure of early-type galaxies at $z\ga0.1$ to their local
counterparts.

\begin{table*}
  \small
  \centering
  \caption{The VIMOS/IFU sample. Apparent brightness $m_{V}$, effective radius
    $R_{\mathrm{eff},V}$ and Einstein radius $\theta_{\mathrm{Einst}}$ are from
    \citet{Auger2009}. Redshifts and SDSS velocity dispersion are from
    \citet{Bolton2008a}. VIMOS velocity dispersions were measured from
    aperture-integrated spectra from the IFU data; errors include only
    the effect of noise, not of template mismatch (5--10\,\%,
    Sect.~\ref{ssec:kinematic:method}). The last two 
    columns specify the grism used for each object and the number of
    observing blocks spent on each target.}
  \begin{tabular}{lcccccccr@{$\pm$}lclr}
    \hline\hline
    Galaxy & $\alpha_\mathrm{J2000}$ & $\delta_\mathrm{J2000}$ & $m_{V}$ & $R_{\mathrm{eff},V}$ & $\theta_\mathrm{Einst}$ & $z_\mathrm{lens}$ & $z_\mathrm{source}$ & \multicolumn{2}{c}{$\sigma_\mathrm{SDSS,B08}$} & $\sigma_{\mathrm{VIMOS}}$ & Grism & OBs \\
           &                        &                        &         & \multicolumn{1}{c}{(arcsec)} & \multicolumn{1}{c}{(arcsec)} &  & & \multicolumn{2}{c}{($\mathrm{km\,s^{-1}}$)} & ($\mathrm{km\,s^{-1}}$) & & \\
    \hline
    SDSS\,J0037  & $00$:$37$:$53.21$ & $-09$:$42$:$20.1$ & $16.90$ & $2.68$ & $1.53$ & $0.1955$ & $0.6322$ & $279$ & $10$ & $245.3^{+ 6.9}_{- 7.2}$ & HR\_Blue   &  9 \\[0.75ex]
    SDSS\,J0216  & $02$:$16$:$52.54$ & $-08$:$13$:$45.3$ & $18.36$ & $2.97$ & $1.16$ & $0.3317$ & $0.5235$ & $333$ & $23$ & $340.7^{+ 7.8}_{- 7.7}$ & HR\_Orange & 14 \\[0.75ex]
    SDSS\,J0912  & $09$:$12$:$05.31$ & $+00$:$29$:$01.2$ & $16.56$ & $4.29$ & $1.63$ & $0.1642$ & $0.3239$ & $326$ & $12$ & $306.5^{+10.9}_{-11.4}$ & HR\_Blue   &  4 \\[0.75ex]
    SDSS\,J0935  & $09$:$35$:$43.93$ & $-00$:$03$:$34.8$ & $17.71$ & $4.12$ & $0.87$ & $0.3475$ & $0.4670$ & $396$ & $35$ & $330.4^{+ 9.0}_{- 8.5}$ & HR\_Orange & 12 \\[0.75ex]
    SDSS\,J0959  & $09$:$59$:$44.07$ & $+04$:$10$:$17.0$ & $17.94$ & $1.51$ & $0.99$ & $0.1260$ & $0.5350$ & $197$ & $13$ & $244.2^{+15.2}_{-14.7}$ & HR\_Orange &  4 \\[0.75ex]
    SDSS\,J1204  & $12$:$04$:$44.07$ & $+03$:$58$:$06.4$ & $17.45$ & $1.65$ & $1.31$ & $0.1644$ & $0.6307$ & $267$ & $17$ & $240.8^{+ 9.3}_{- 9.5}$ & HR\_Orange &  5 \\[0.75ex]
    SDSS\,J1250A & $12$:$50$:$28.26$ & $+05$:$23$:$49.1$ & $17.77$ & $1.91$ & $1.13$ & $0.2318$ & $0.7953$ & $252$ & $14$ & $219.9^{+ 5.7}_{- 5.4}$ & HR\_Orange &  6 \\[0.75ex]
    SDSS\,J1250B & $12$:$50$:$50.52$ & $-01$:$35$:$31.7$ & $15.68$ & $4.01$ &     -- & $0.0871$ & $0.3529$ & $246$ & $ 9$ & $274.4^{+ 6.9}_{- 6.7}$ & HR\_Orange &  6 \\[0.75ex]
    SDSS\,J1251  & $12$:$51$:$35.71$ & $-02$:$08$:$05.2$ & $17.71$ & $5.34$ & $0.84$ & $0.2243$ & $0.7843$ & $233$ & $23$ & $201.5^{+ 8.1}_{- 8.7}$ & HR\_Orange & 12 \\[0.75ex]
    SDSS\,J1330  & $13$:$30$:$45.53$ & $-01$:$48$:$41.6$ & $17.56$ & $1.36$ & $0.87$ & $0.0808$ & $0.7115$ & $185$ & $ 9$ & $191.8^{+ 7.8}_{- 7.5}$ & HR\_Orange &  3 \\[0.75ex]
    SDSS\,J1443  & $14$:$43$:$19.62$ & $+03$:$04$:$08.2$ & $17.62$ & $1.38$ & $0.81$ & $0.1338$ & $0.4187$ & $209$ & $11$ & $226.8^{+ 9.7}_{- 9.2}$ & HR\_Orange &  4 \\[0.75ex]
    SDSS\,J1451  & $14$:$51$:$28.19$ & $-02$:$39$:$36.4$ & $16.92$ & $2.64$ & $1.04$ & $0.1254$ & $0.5203$ & $223$ & $14$ & $203.9^{+ 5.4}_{- 5.3}$ & HR\_Orange &  7 \\[0.75ex]
    SDSS\,J1627  & $16$:$27$:$46.45$ & $-00$:$53$:$57.6$ & $17.87$ & $2.05$ & $1.23$ & $0.2076$ & $0.5241$ & $290$ & $14$ & $272.6^{+ 7.8}_{- 8.9}$ & HR\_Orange & 11 \\[0.75ex]
    SDSS\,J2238  & $22$:$38$:$40.20$ & $-07$:$54$:$56.0$ & $17.18$ & $2.41$ & $1.27$ & $0.1371$ & $0.7126$ & $198$ & $11$ & $226.8^{+ 6.2}_{- 6.2}$ & HR\_Orange &  6 \\[0.75ex]
    SDSS\,J2300  & $23$:$00$:$53.15$ & $+00$:$22$:$38.0$ & $18.19$ & $1.93$ & $1.24$ & $0.2285$ & $0.4635$ & $279$ & $17$ & $250.4^{+ 6.4}_{- 7.0}$ & HR\_Orange & 10 \\[0.75ex]
    SDSS\,J2303  & $23$:$03$:$21.72$ & $+14$:$22$:$17.9$ & $16.77$ & $3.54$ & $1.62$ & $0.1553$ & $0.5170$ & $255$ & $16$ & $301.5^{+ 7.1}_{- 7.1}$ & HR\_Orange & 11 \\[0.75ex]
    SDSS\,J2321  & $23$:$21$:$20.93$ & $-09$:$39$:$10.3$ & $15.27$ & $4.79$ & $1.60$ & $0.0819$ & $0.5324$ & $249$ & $ 8$ & $248.1^{+ 4.4}_{- 4.4}$ & HR\_Blue   &  5 \\[0.75ex]
    \hline
  \end{tabular}
  \label{tab:sample}
\end{table*}

\section{Observations}
\label{sec:Observations}

\subsection{Integral-field spectroscopy}
\label{ssec:obs:VIMOS}

We have obtained integral-field spectroscopic observations of
17~systems using the integral-field unit (IFU) of VIMOS
\citep{LeFevre2001short}.

VIMOS is a wide-field imager, multi-object and integral-field
spectrograph mounted at the Nasmyth focus~B of the Very Large
Telescope UT3 (Melipal) on Paranal, Chile. The integral-field unit
samples the focal plane with a total of 6400~lenslets, each of which
is coupled to an optical fibre. The opposite ends of the fibres are
arranged to form pseudo-slits of 400 fibres each. The light emerging
from the pseudo-slit is dispersed in the usual way using grisms and
optionally order-separating filters, and the resulting spectra are
recorded on four $2\mathrm{k}\times4\mathrm{k}$ EEV CCD chips. In
low-resolution mode, the spectra from four pseudo-slits are stacked in
the dispersion direction on each of the four CCDs. The arrangement is
such that each CCD records the spectra from one quadrant of the field
of the IFU head. In high-resolution mode, only the spectra from one
pseudo-slit fit onto a CCD and only the central 1600 elements of the
IFU head are used. The spatial scale in the focal plane can be chosen
to be $0.67\,\mathrm{arcsec}$ or $0.33\,\mathrm{arcsec}$ per spectral
element. We used the large spatial scale and high-resolution mode and
thus work with a field of view of $27\arcsec\times27\arcsec$ covered
by $40\times40$ spectra.

Three of the 17 systems of the sample were observed as ESO Programme
075.B-0226 (henceforth the ``pilot programme'') between June 2005 and
January 2006. The remaining fourteen systems were the targets of an
ESO Large Programme, 177.B-0682 (the ``main programme''), and were
observed between April 2006 and March 2007. For the pilot programme,
the high-resolution blue grism (HR-Blue) was used, which has a nominal
resolution of $R=2550$ and covers the wavelength range $4000\,$\AA\ to
6200\,\AA; for the main programme the HR-Orange grism was used, which
covers a somewhat redder wavelength range between 5000\,\AA\ and
7000\,\AA\ at comparable resolution ($R=2650$).

All observations were carried out in service mode. For this, the total
observing time was broken down in observing blocks (OBs), each with an
execution time of one hour and comprising science and calibration
exposures. For the pilot programme, three dithered science exposures
of integration time 555~seconds on one target were taken during an
OB. For the main programme, a single exposure per OB was taken, with
integration time 2060~seconds. The main observational constraint for
scheduling the OBs for this project was a maximum seeing of
$0.8\,\mathrm{arcsec}$ FWHM (full width at half maximum).

We followed ESO's standard calibration plan which comprises three
flat-field exposures and one arc-lamp exposure per OB, taken
immediately after the science exposures. Bias exposures were taken
during day time in blocks of five exposures.

Standard stars are observed at varying intervals by ESO staff and were
therefore in most cases not available for the same nights that our
science data were taken. They are useful for relative flux calibration
but not for absolute spectro-photometric calibration.

\subsection{HST imaging}
\label{ssec:obs:HST}

Modelling stellar dynamics and the lens effect requires photometric
information on both the lens galaxy and the lensed background
source. We use various HST observations obtained for the SLACS project
\citep[see][]{Bolton2008a, Auger2009}.

Lens modelling aims at reconstructing the two-dimensional
surface-brightness distribution of the background source. This is
obtained from high-resolution ACS images from which elliptical
B-spline models of the lens galaxies were subtracted \citep[see][for
details]{Bolton2008a}. For eight systems, we have deep F814W images
from Programmes 10494 and 10798 (Cycle 14, PI: L.~Koopmans) consisting
of four exposures. For the remaining systems only single exposure
images from snapshot Programme 10174 (Cycles 13, PI: L.~Koopmans) are
available. The pixel scale of these images is $0.05\,\mathrm{arcsec}$.

The surface-brightness distributions of the lens galaxies are obtained
from NICMOS F160W images from Programmes 10494, 10798 and 11202 (Cycle
16, PI: Koopmans). The full resolution images have a pixel scale of
$0.05\,\mathrm{arcsec}$ and include four exposures for a total
exposure time of $2560\,\mathrm{s}$. In order to have proper flux
weights for the kinematic maps we resample the NICMOS images to the
VIMOS/IFU grid with $0.67\,\mathrm{arcsec}$ per spatial pixel after
convolution to match the point spread function (PSF) of the VIMOS data,
taken to be Gaussian with FWHM of $0.8\,\mathrm{arcsec}$.

Three systems (J1251, J1627 and J2321) were not observed with
NICMOS. For these systems we use the B-spline models to the F814W ACS
images to estimate the surface brightness of the lens galaxy. We add
random noise according to the ACS weight maps to mimic observational
data.

All images were astrometrically registered using the measured position
of the lens galaxy and in some cases additional objects visible in the
VIMOS field of view.

\begin{figure*}
  \centering
  \resizebox{0.22\hsize}{!}{\includegraphics[clip]{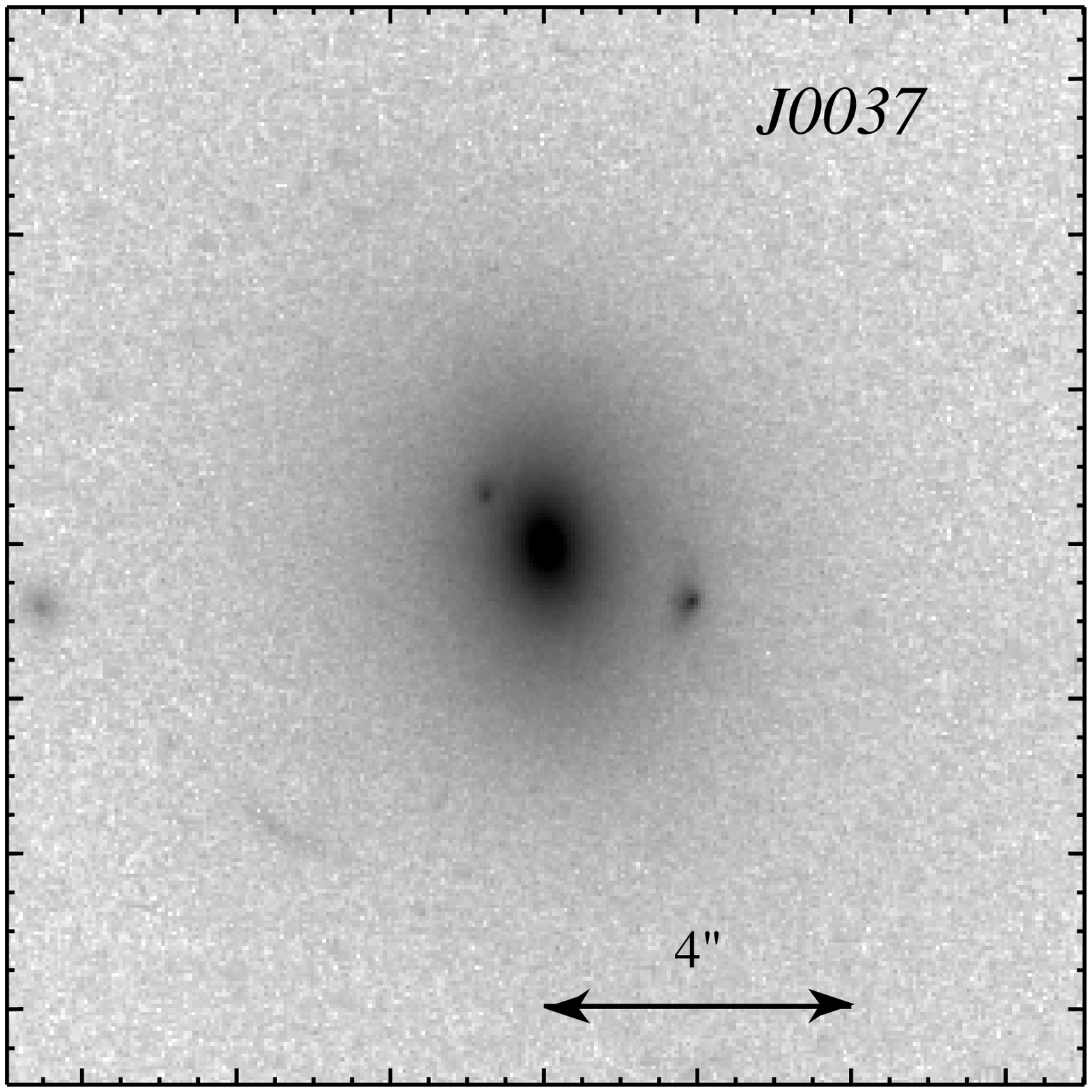}}\hfill
  \resizebox{0.22\hsize}{!}{\includegraphics[clip]{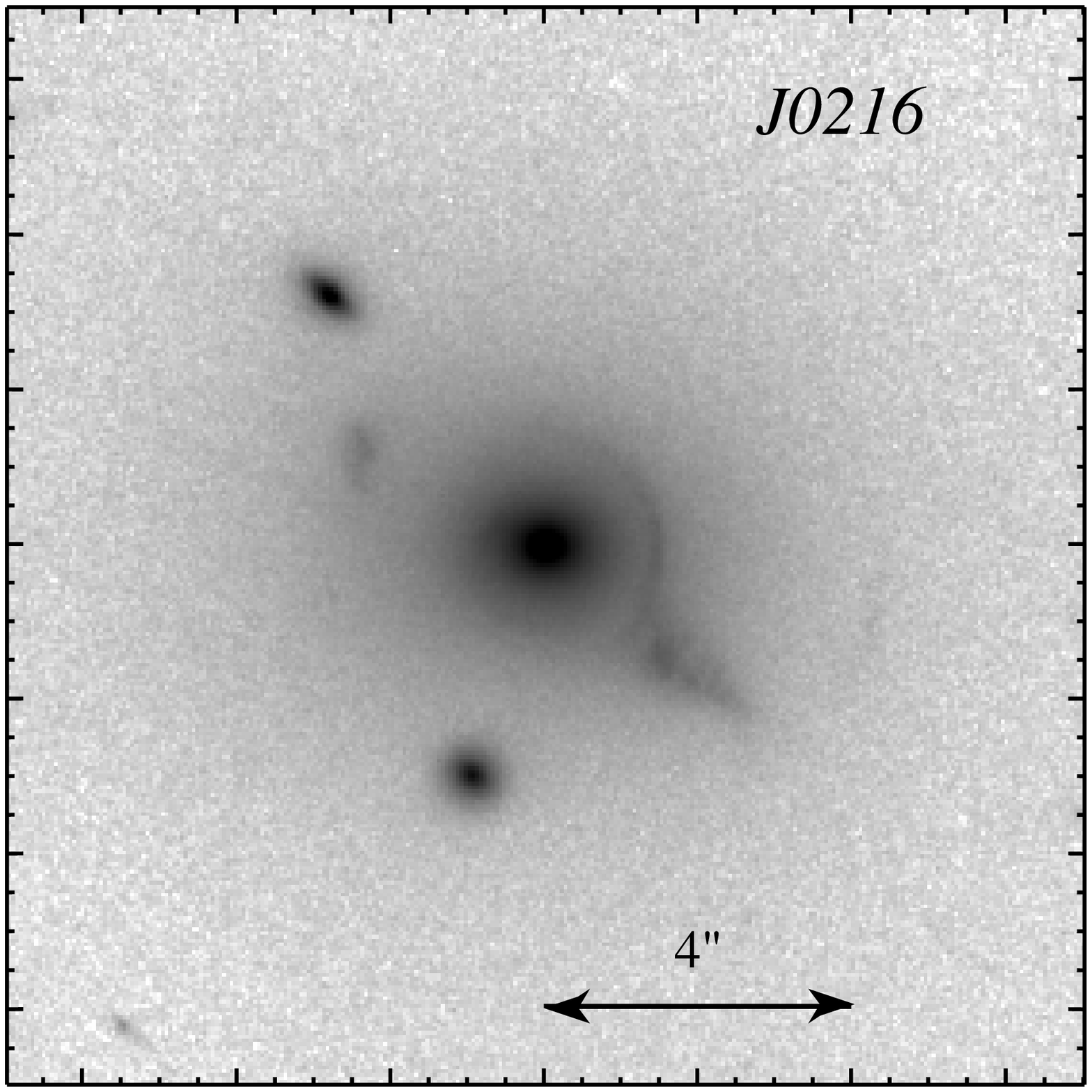}} \hfill
  \resizebox{0.22\hsize}{!}{\includegraphics[clip]{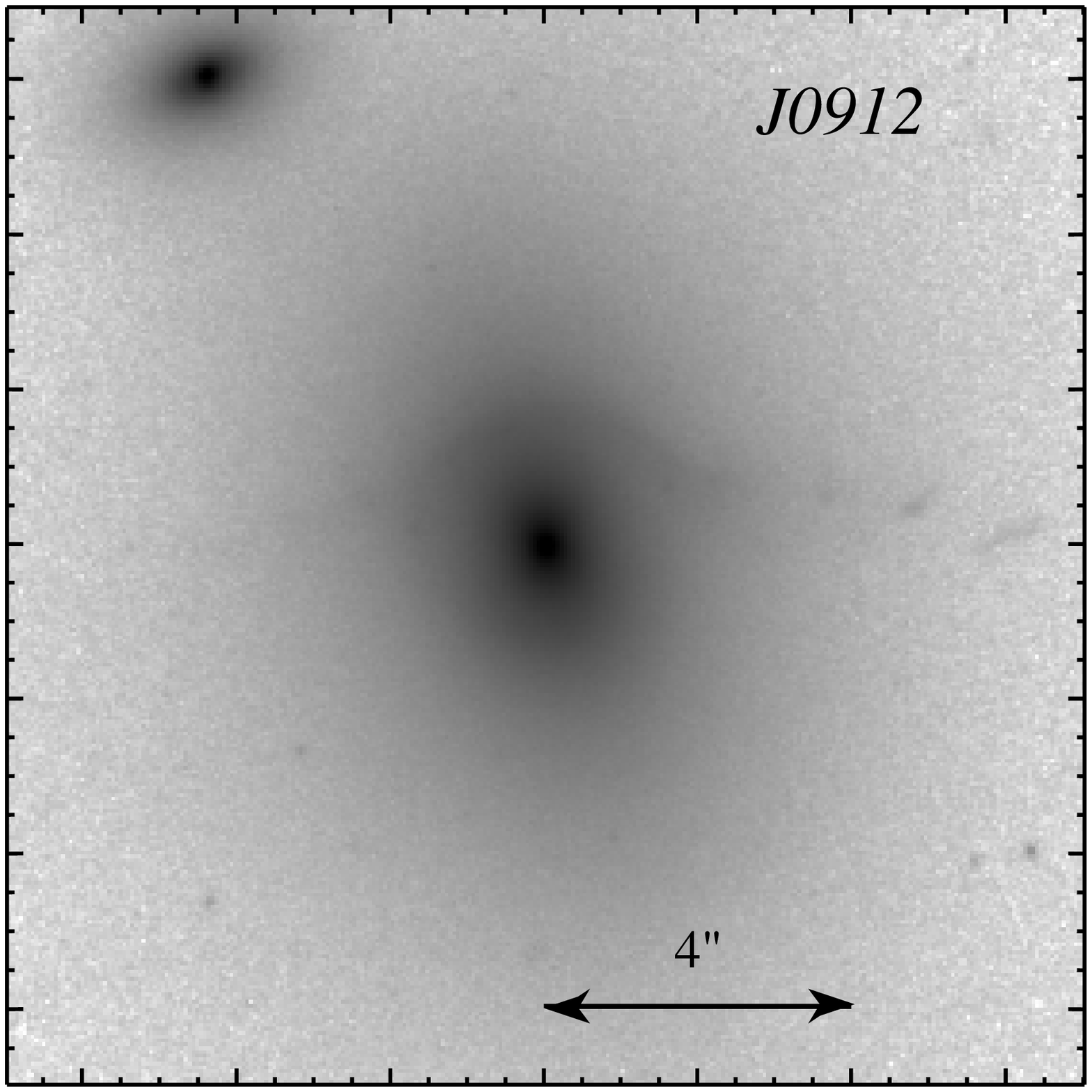}}\hfill
  \resizebox{0.22\hsize}{!}{\includegraphics[clip]{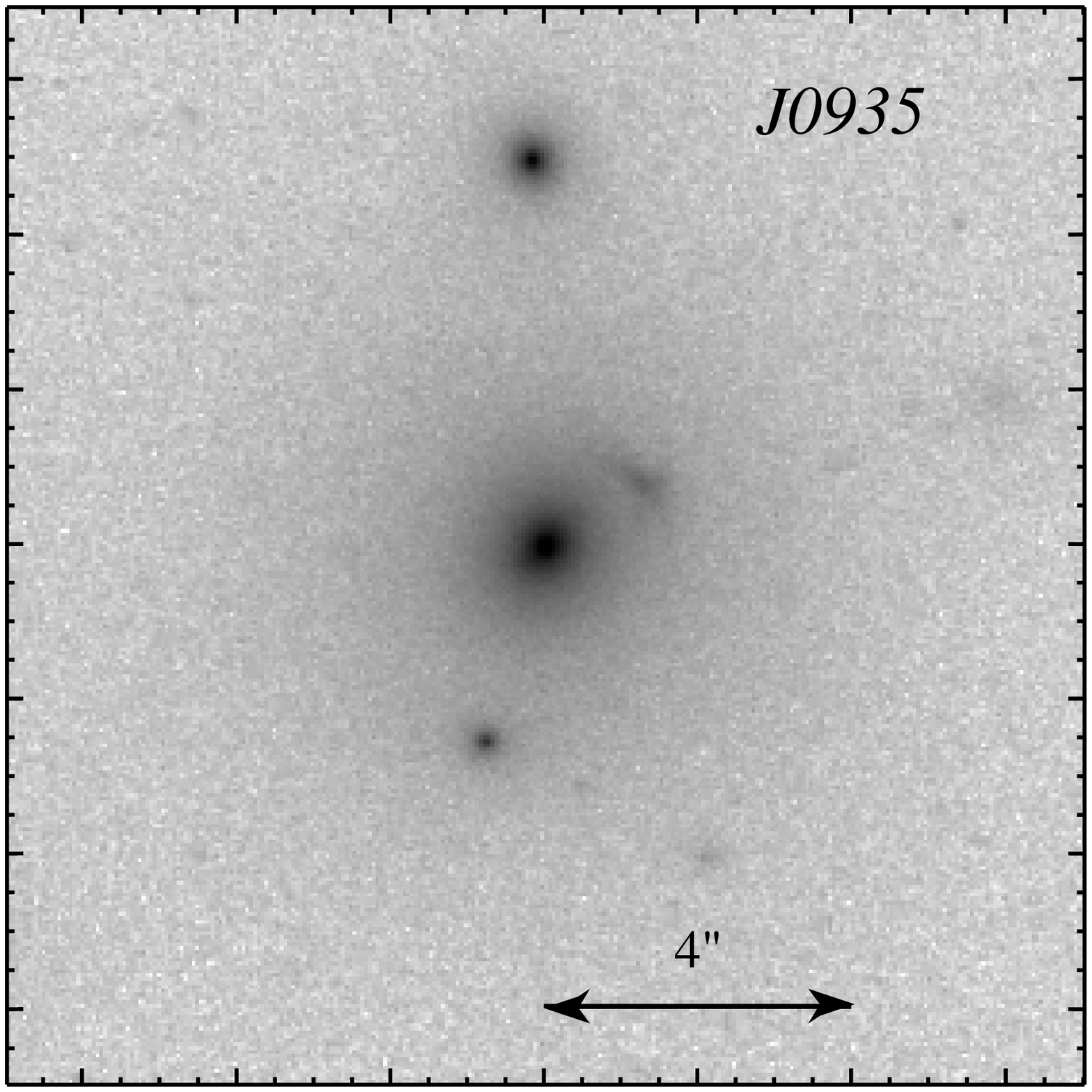}}\\[2ex]
  \resizebox{0.22\hsize}{!}{\includegraphics[clip]{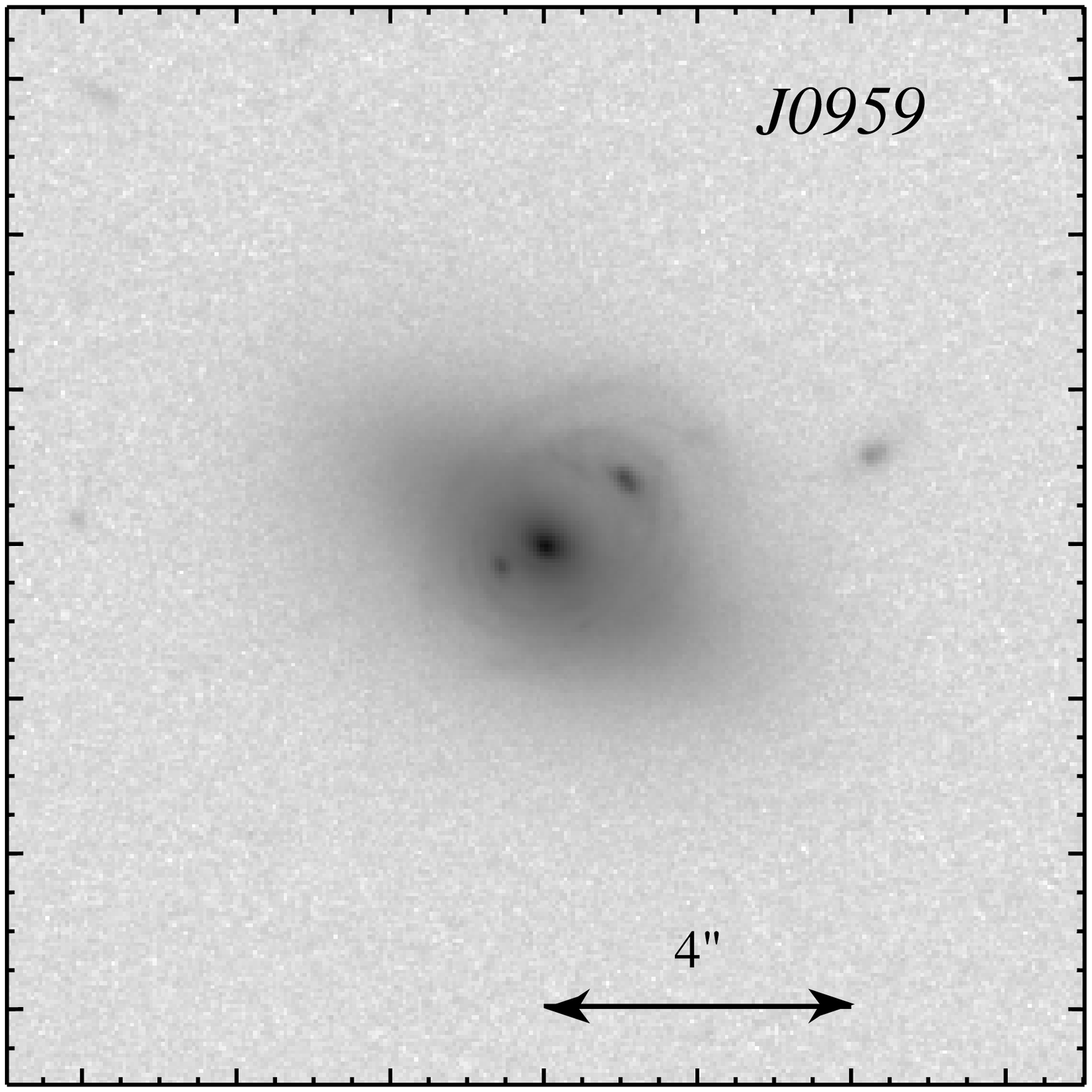}}\hfill  
  \resizebox{0.22\hsize}{!}{\includegraphics[clip]{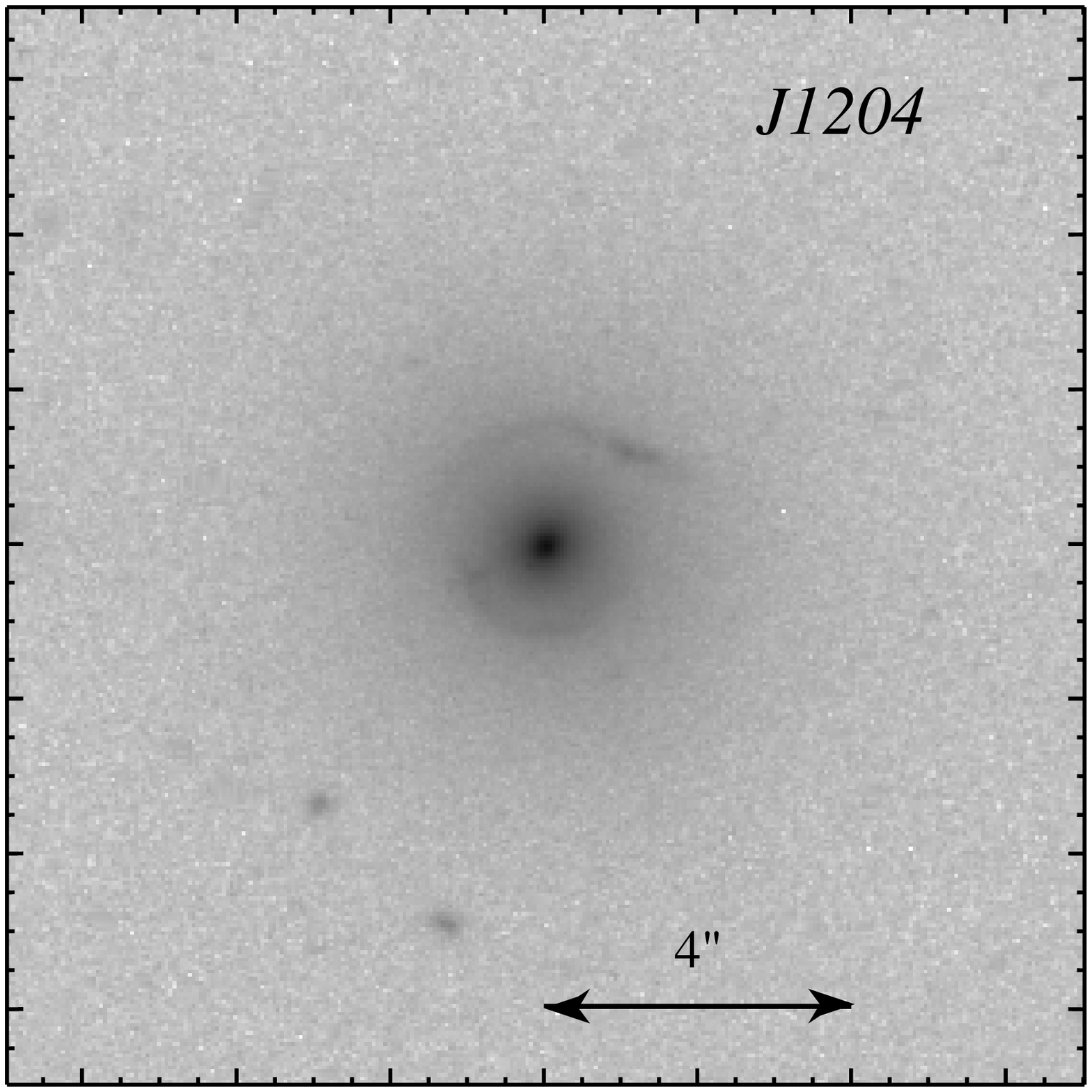}}\hfill
  \resizebox{0.22\hsize}{!}{\includegraphics[clip]{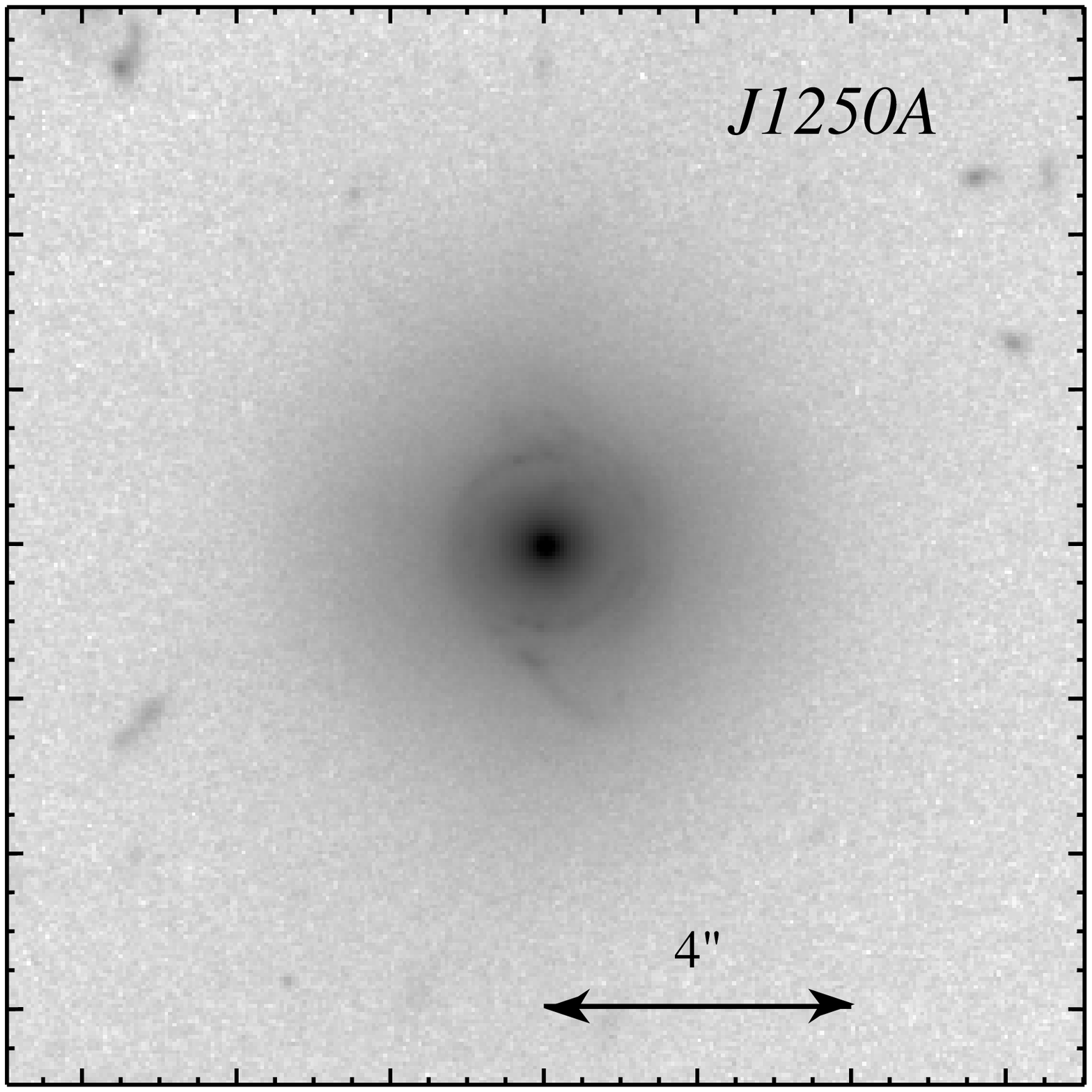}}\hfill  
  \resizebox{0.22\hsize}{!}{\includegraphics[clip]{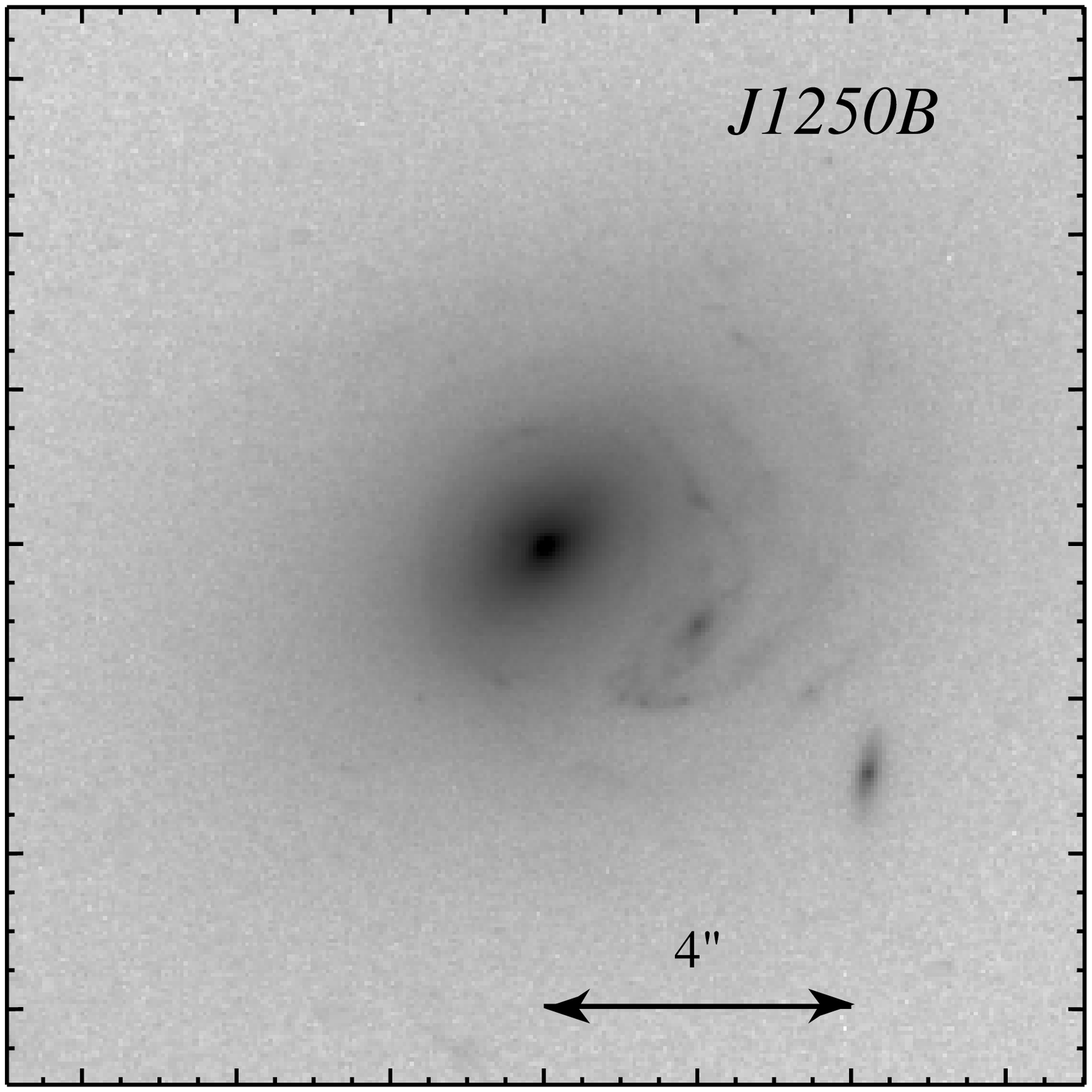}}\\[2ex]
  \resizebox{0.22\hsize}{!}{\includegraphics[clip]{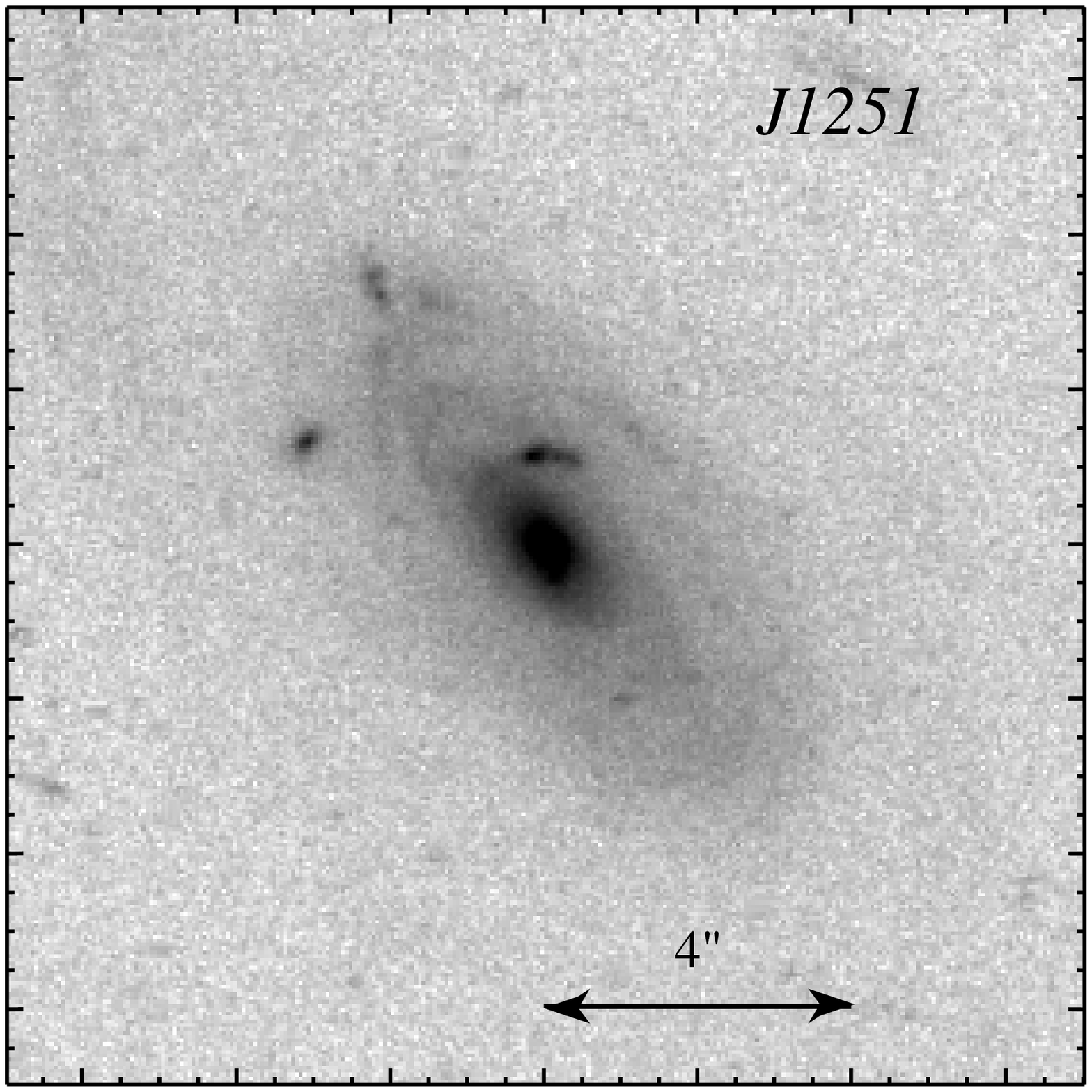}}\hfill
  \resizebox{0.22\hsize}{!}{\includegraphics[clip]{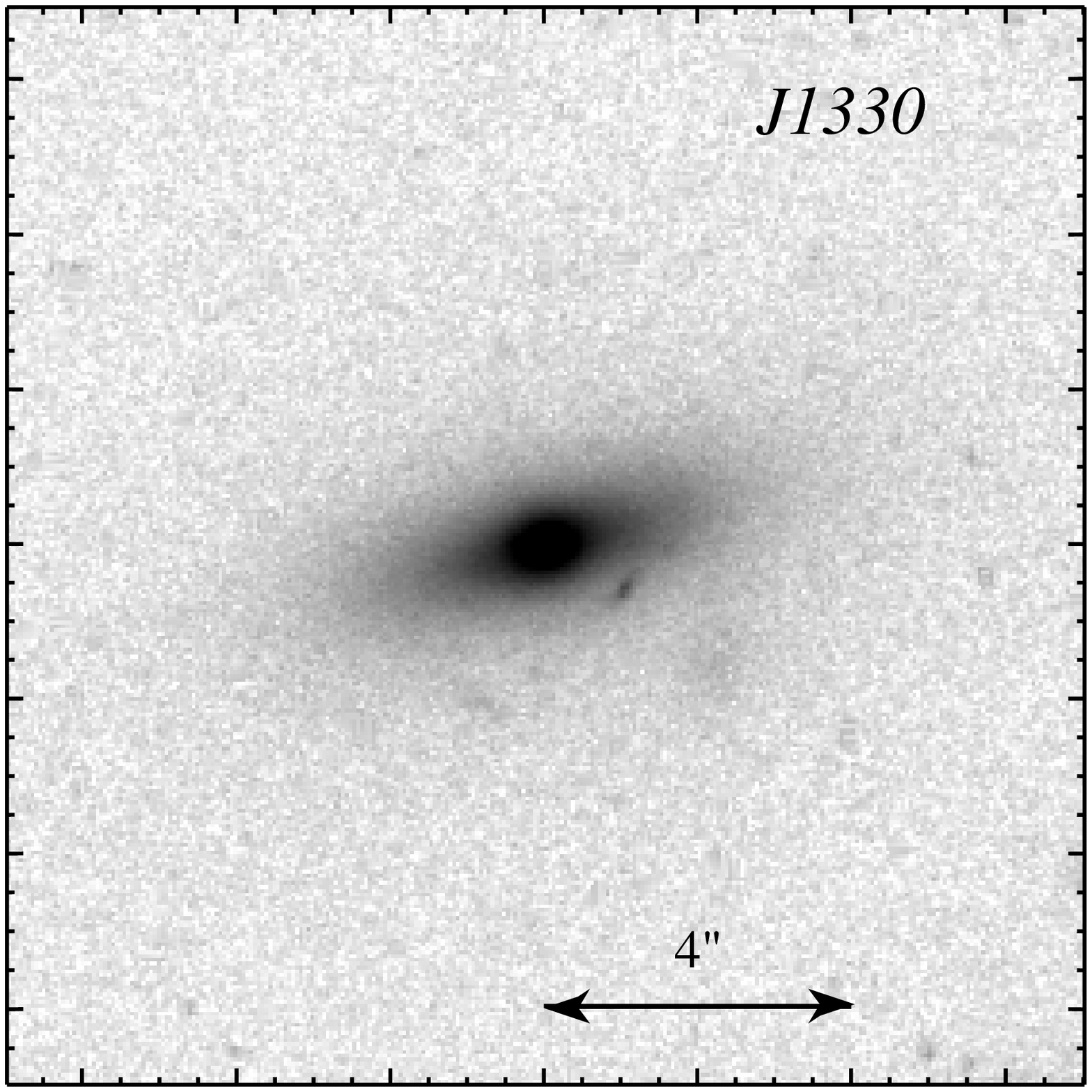}}\hfill
  \resizebox{0.22\hsize}{!}{\includegraphics[clip]{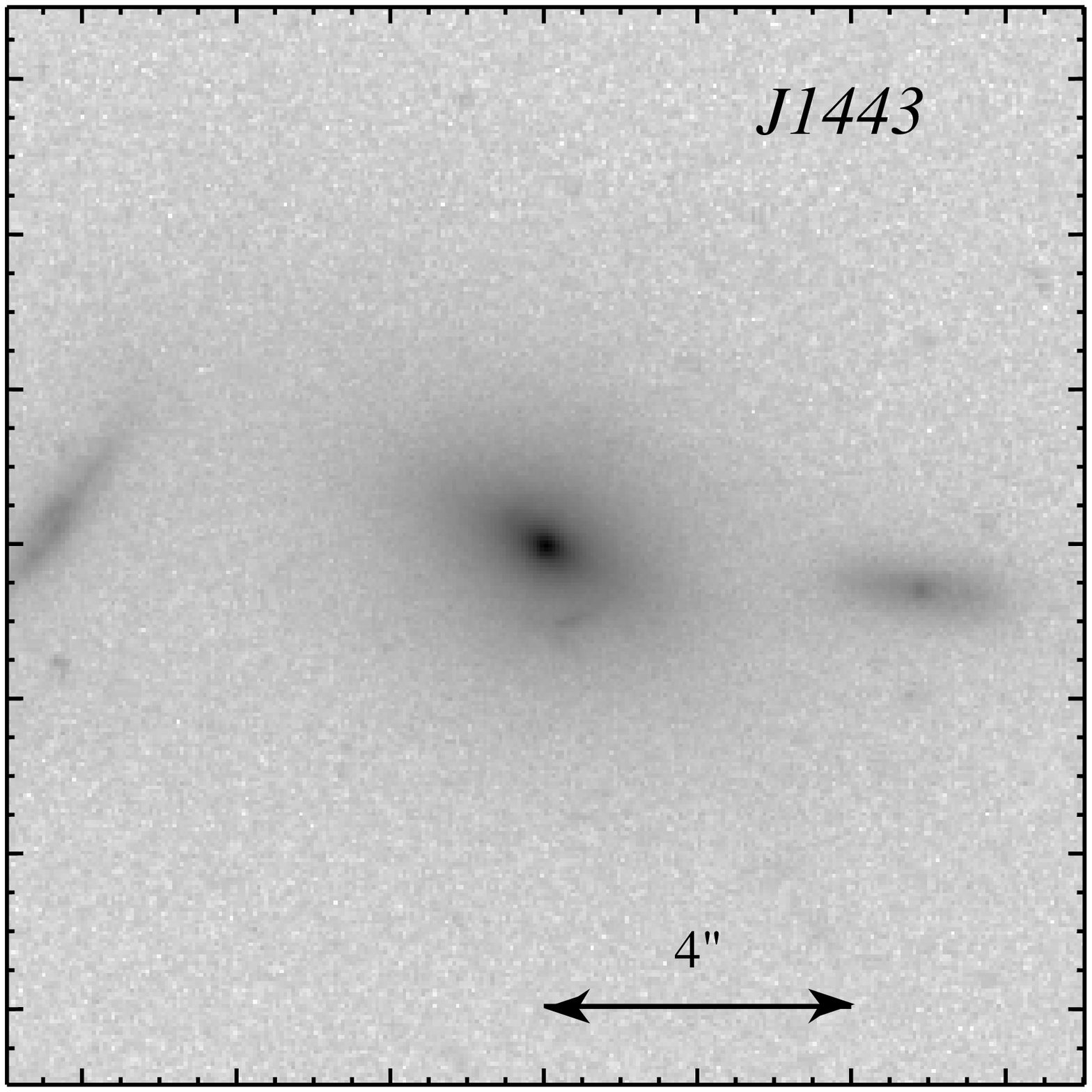}}\hfill
  \resizebox{0.22\hsize}{!}{\includegraphics[clip]{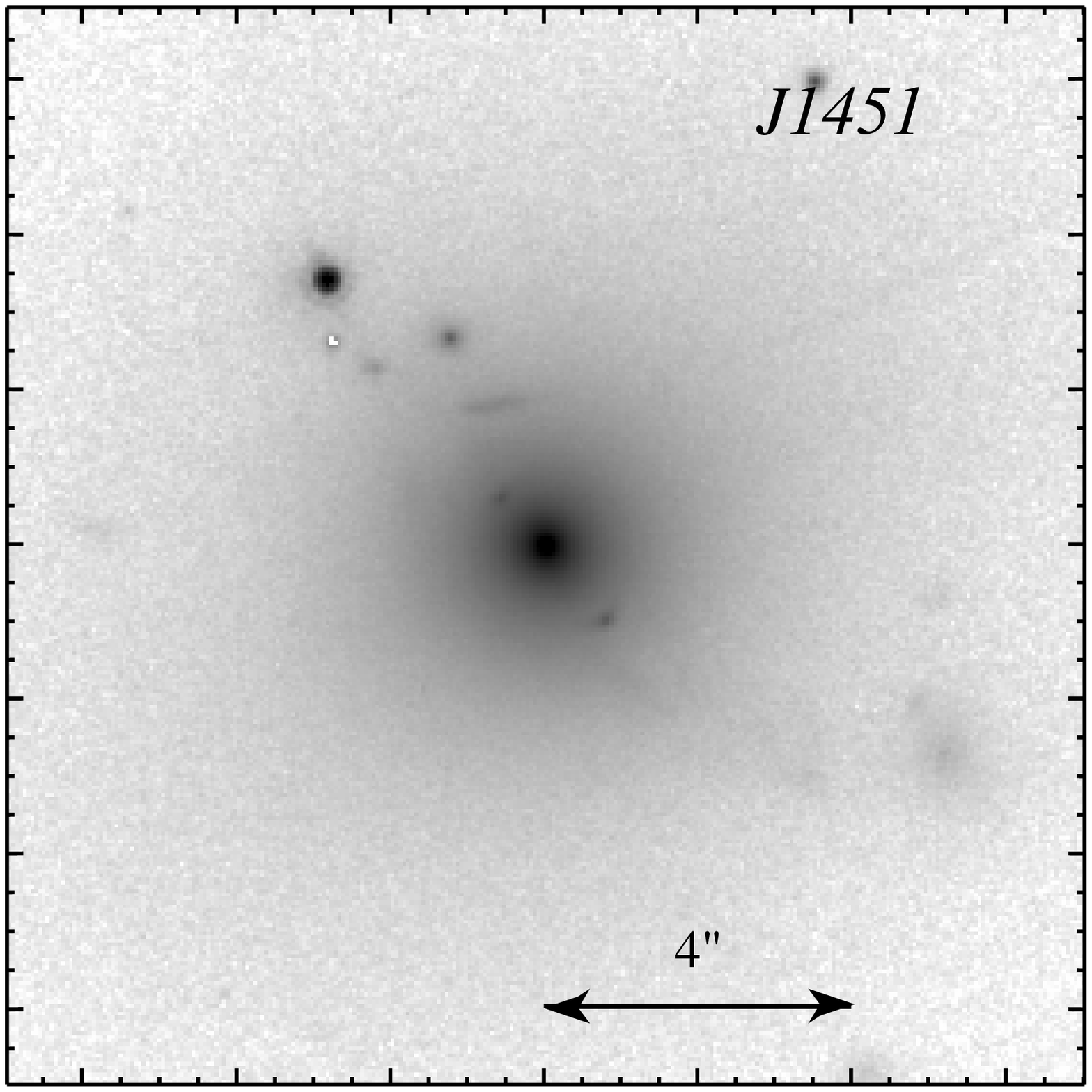}}\\[2ex]
  \resizebox{0.22\hsize}{!}{\includegraphics[clip]{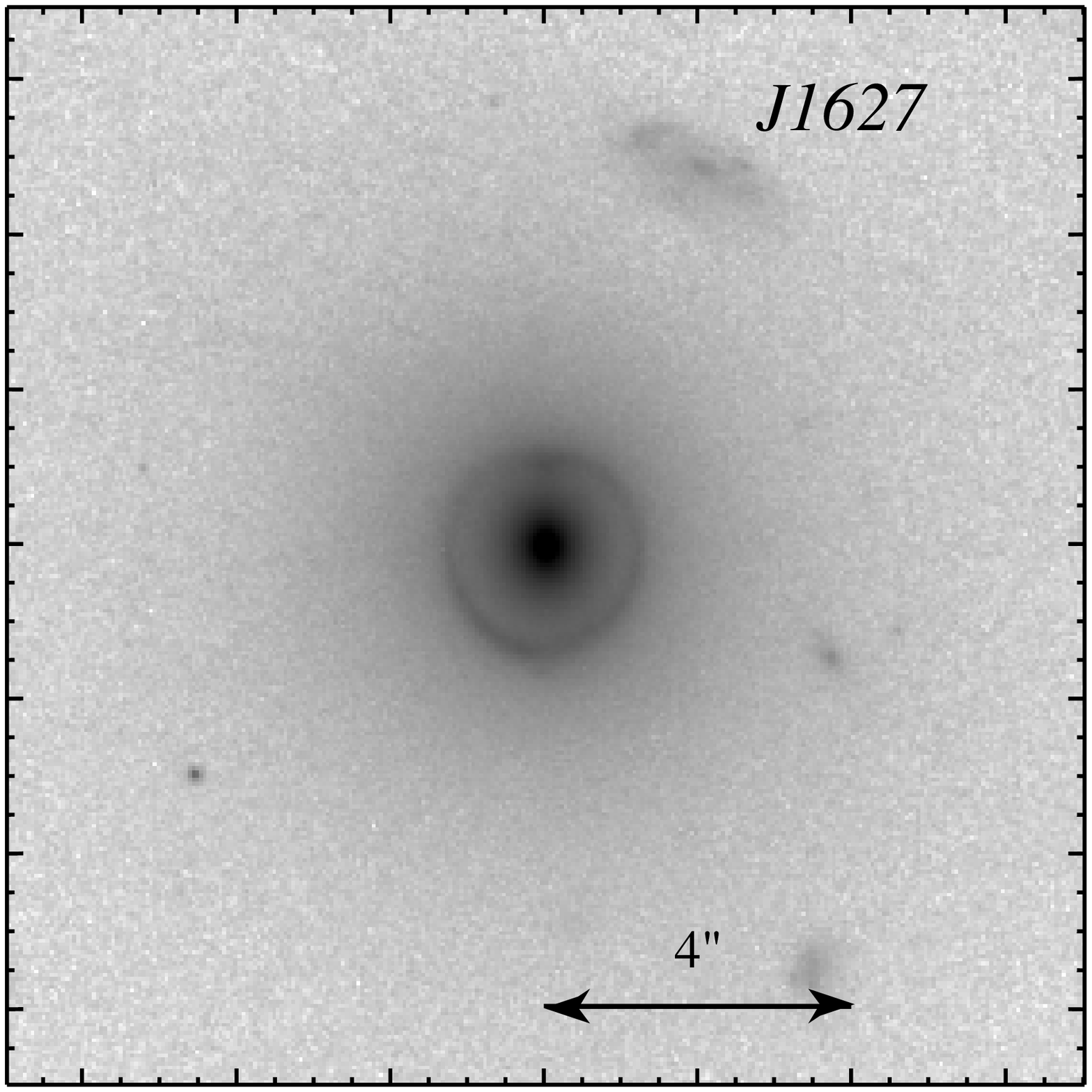}}\hfill  
  \resizebox{0.22\hsize}{!}{\includegraphics[clip]{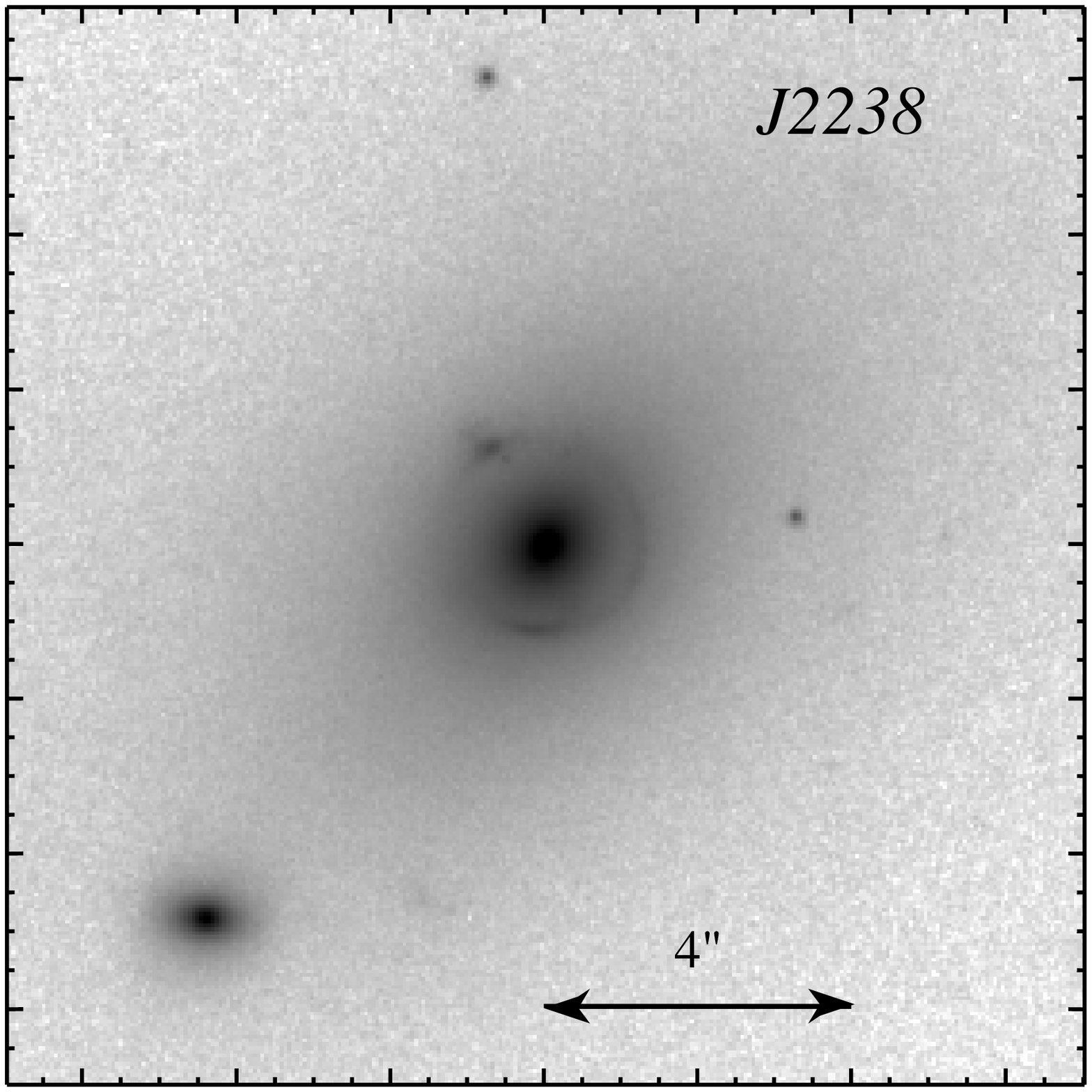}}\hfill
  \resizebox{0.22\hsize}{!}{\includegraphics[clip]{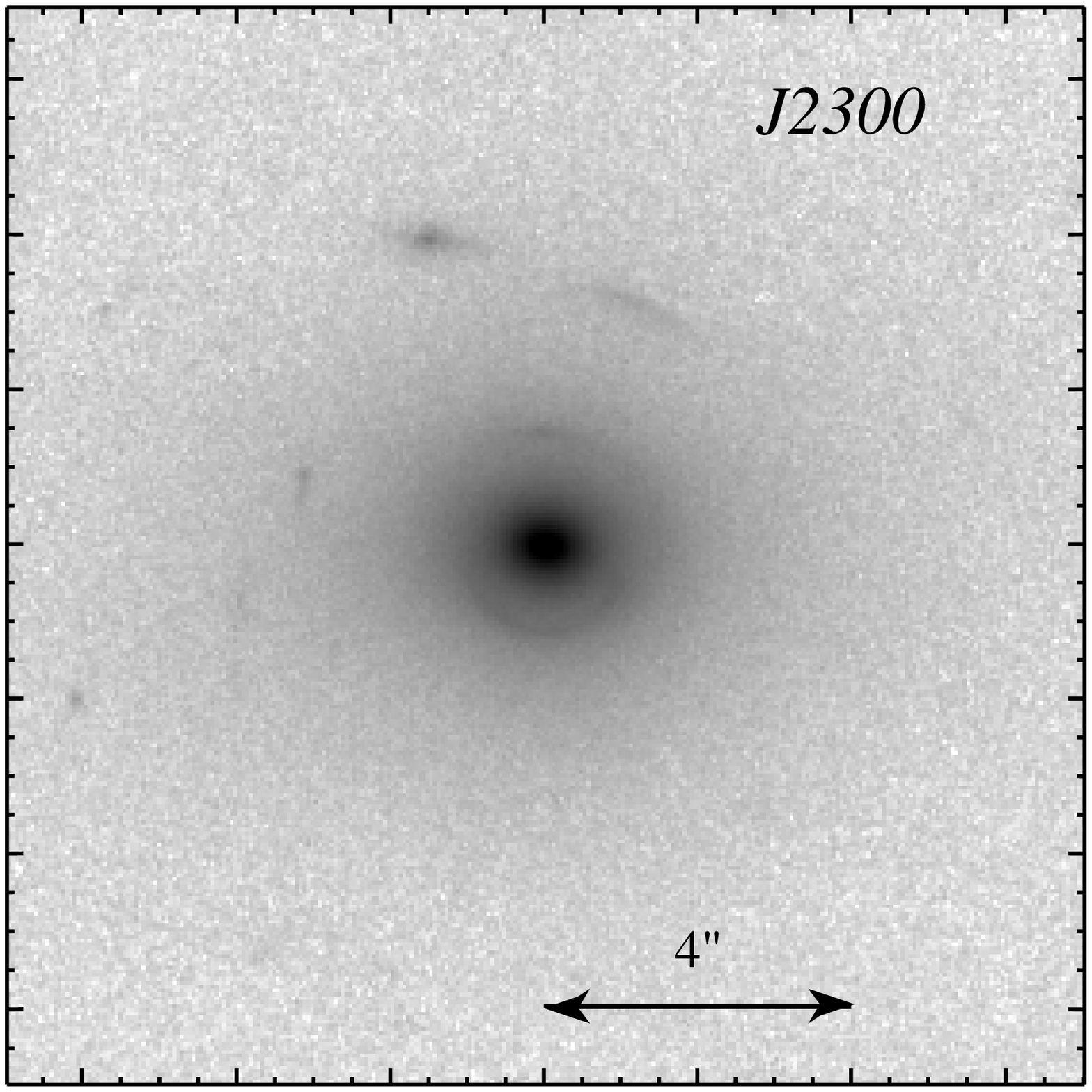}}\hfill
  \resizebox{0.22\hsize}{!}{\includegraphics[clip]{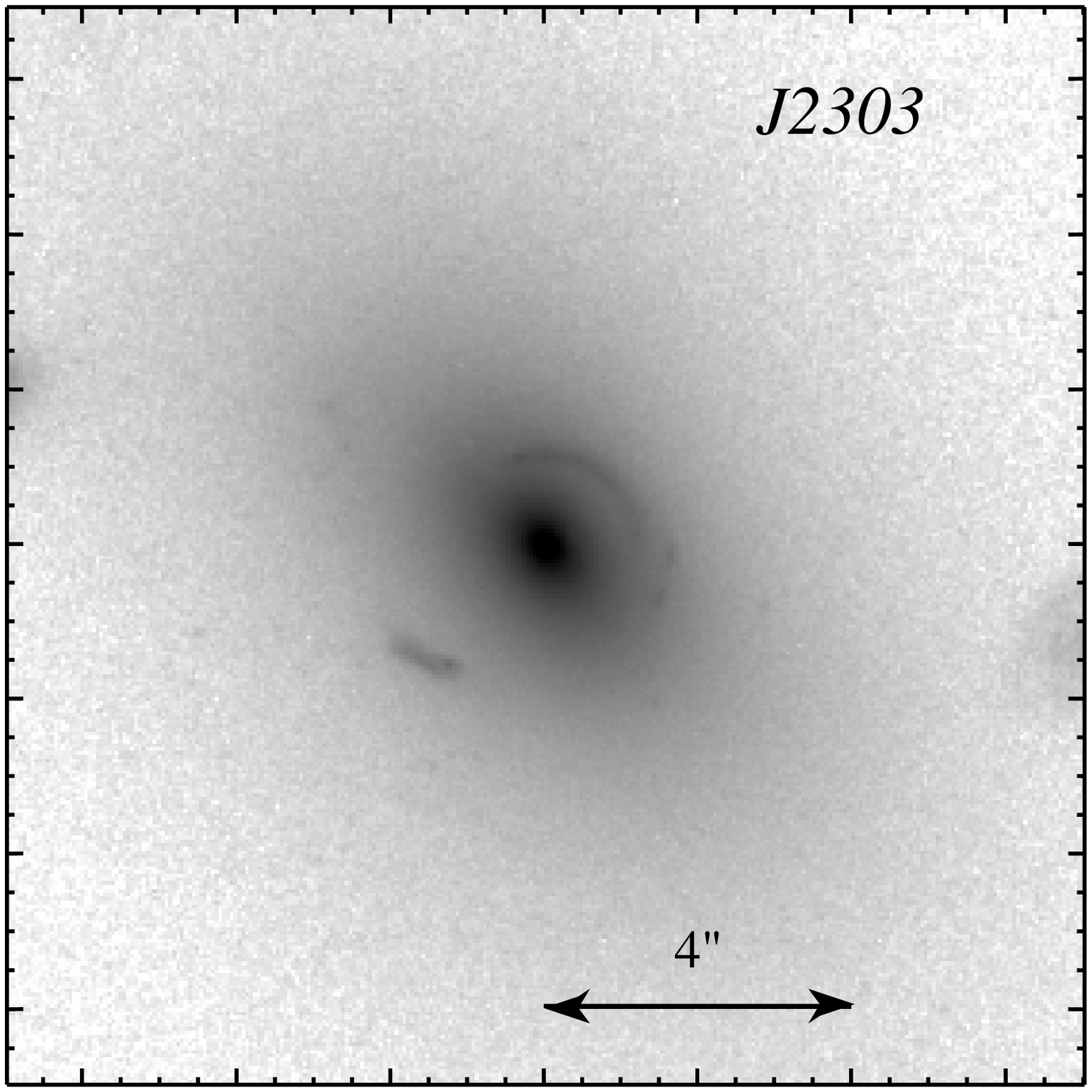}}\\[2ex]
  \resizebox{0.22\hsize}{!}{\includegraphics[clip]{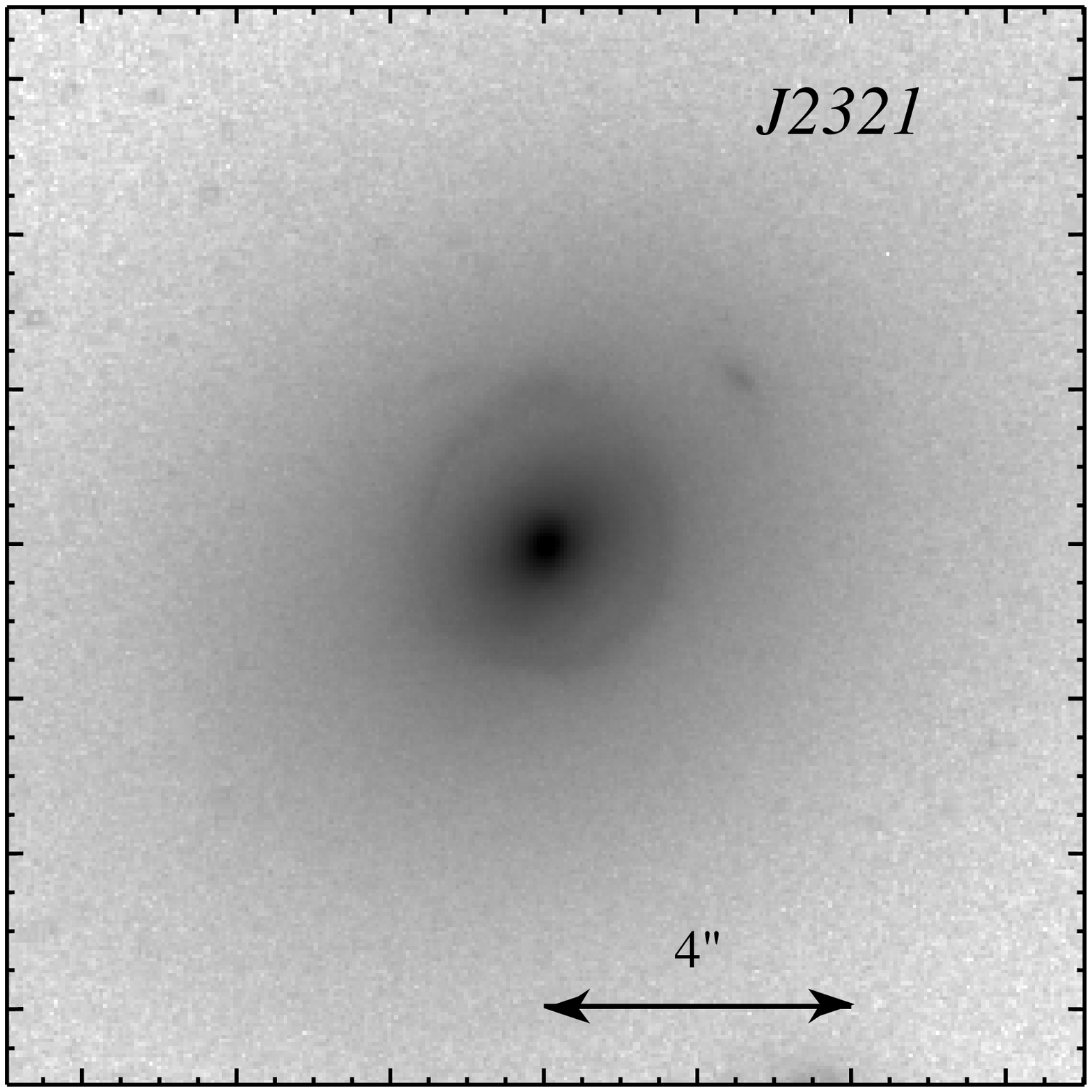}}
  \caption{HST/F814W images of the systems in the sample. In images
    obtained in a snapshot programme cosmic ray hits were interpolated
    over for display purposes.}
  \label{fig:HST_Images}
\end{figure*}

The F814W images of the systems in the sample are shown in
Fig.~\ref{fig:HST_Images}.

\section{Data reduction}
\label{sec:data_reduction}

The VLT VIMOS/IFU data were reduced using the \texttt{vipgi} package
which was developed within the framework of the VIRMOS consortium and
the VVDS project.
\texttt{vipgi}\footnote{http://cosmos.iasf-milano.inaf.it/pandora/vipgi.html}
has been described in detail by \citet{Scodeggio2005short} and
\citet{Zanichelli2005short}.  In \citet{Czoske2008}, we have already
given an overview of the data reduction procedure as applied to our
data set, so that we restrict ourselves here to giving the main
parameters.

The wavelength calibration is based on about 20 helium and neon lines
spread over the wavelength range. Fitting a third-order polynomial
results in residuals with root-mean-square of $\sim 0.075\,$\AA,
corresponding to $\sim 5\,\mathrm{km\,s^{-1}}$, significantly smaller
than the spectral resolution and negligible in our analysis.

After rectification of the two-dimensional spectra onto a linear
wavelength grid with a dispersion of $0.644$\,\AA\ per pixel,
one-dimensional spectra are extracted for each fiber in an optimal way
using the algorithm of \citet{Horne1986}. We refer to the resulting
collection of one-dimensional spectra as a ``data cube'' even though
the data structure is, strictly speaking, not a three-dimensional
cube. Spectra are labeled by the spatial coordinates $L$ and $M$ (or
world coordinates $\alpha$ and $\delta$).

A relative flux calibration (giving correct flux as a function of
wavelength) is done using the standard observation nearest in time to
the execution of a given science observing block. Since this does not
necessarily come from the same night as the science exposures and no
attempt is made to observe them at the same airmass, the standard
observations can only provide a relative correction but no absolute
spectro-photometric calibration.

The varying transmissivity of the fibres is corrected by measuring and
comparing the flux under a sky line in the different extracted
spectra. We used [\ion{O}{i}]\,$\lambda 5577.4$ for this purpose,
integrating over a range of seven pixels ($4.2~$\AA) and subtracting
an estimate of the underlying continuum from a window of five pixels
on one side of the line.

The observed galaxies are smaller than the field of view of VIMOS/IFU,
leaving a sufficient number of fibres pointing at blank sky to permit
good sky subtraction. These sky fibres are identified automatically
from a histogram of the total fluxes recorded in each fibre,
integrated over the entire wavelength range. The sky fibres are then
grouped according to the shape of the [\ion{O}{i}]\,$\lambda 5577$
line, a mean over the sky spectra in each group is computed and
subtracted from each row according to the group it is judged to belong
to.  The sky subtraction works well even redwards of 6200\,\AA, where
the sky is dominated by the OH molecular bands.

The data cubes from all the exposures on a field are finally combined
into a final data cube by taking the median. Telescope offsets between
the exposures are corrected through spatial shifting by an integer
number of fibres. Subpixel shifts could in principle be corrected for
by interpolation between adjacent fibres but in practice it turns out
that the centroids of the galaxies in image reconstructions of the
individual exposures are hard to measure to subpixel accuracy, given
the level of accuracy of the fibre transmission correction and the
resulting noise in the image reconstructions.

A short-coming of \textsc{vipgi} is that it does not provide noise
estimates on the reduced spectra. Since proper weighting is desirable
for the determination of kinematic and stellar population parameters,
we reconstruct noise spectra from a simple model including photon
noise and read-out noise.  We require noise spectra for each spectral
element in the data cube and for the aperture-summed spectra. It turns
out that at each step of the reduction procedure, the noise estimate
can be written as a rescaled version of the current intermediate data
product, including the sky background. The necessary noise spectra can
thus easily be produced by creating and rescaling a second data set,
reduced in the normal manner but with sky subtraction turned off. A
detailed derivation of the noise estimation procedure is given in
Appendix~\ref{sec:noise_estimation}, which can also be read as a
walk-through of the data reduction process.

\begin{figure*}
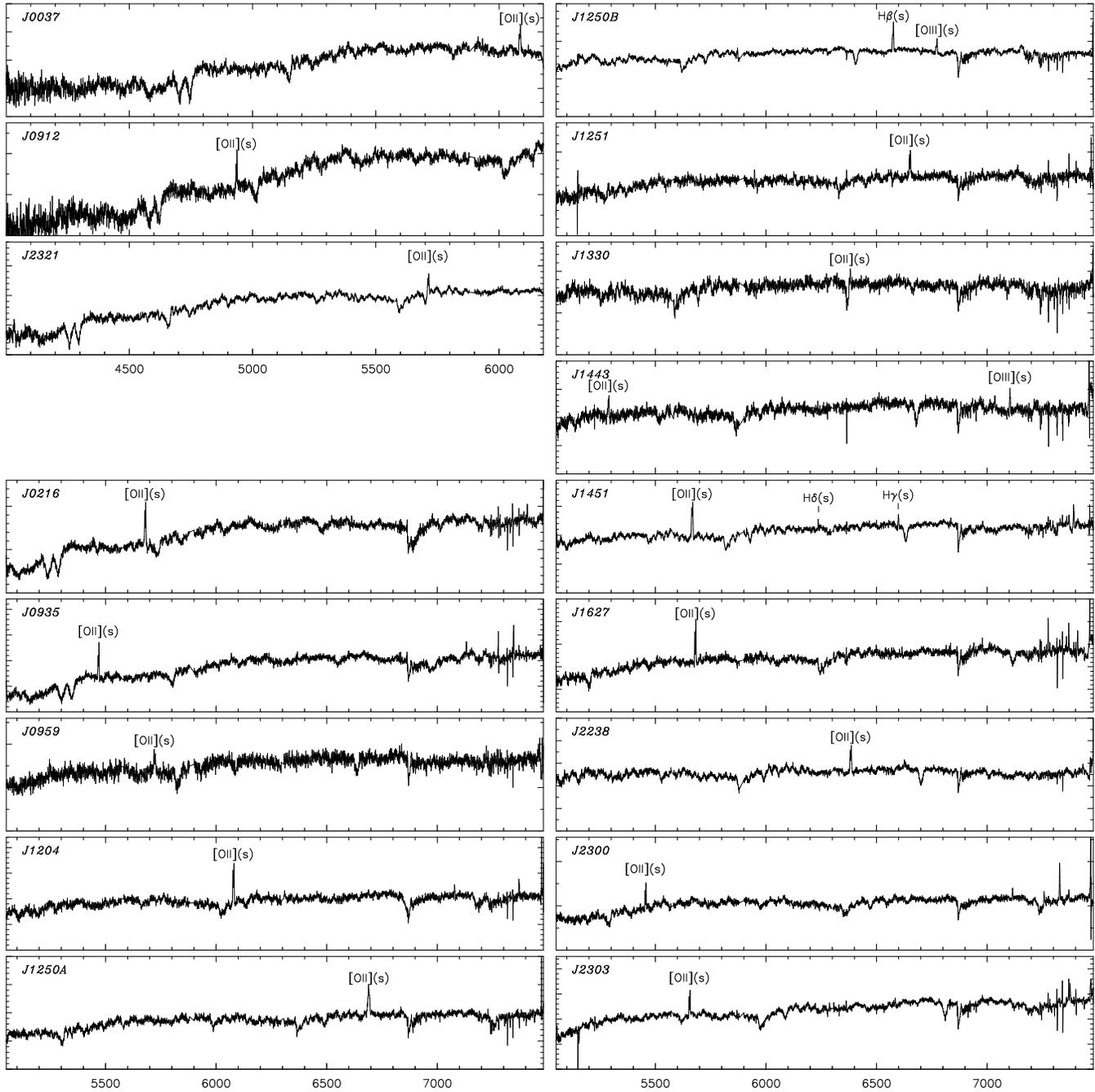

  \centering
  \resizebox{0.495\hsize}{!}{\includegraphics{F03-J0037-spec}} \hfill\resizebox{0.495\hsize}{!}{\includegraphics{F03-J1250B-spec}}\\[-2ex]
  \resizebox{0.495\hsize}{!}{\includegraphics{F03-J0912-spec}} \hfill\resizebox{0.495\hsize}{!}{\includegraphics{F03-J1251-spec}} \\[-2ex]
  \resizebox{0.495\hsize}{!}{\includegraphics{F03-J2321-spec}} \hfill\resizebox{0.495\hsize}{!}{\includegraphics{F03-J1330-spec}} \\[-2ex]
\                                                              \hfill\resizebox{0.495\hsize}{!}{\includegraphics{F03-J1443-spec}} \\[-2ex]
  \resizebox{0.495\hsize}{!}{\includegraphics{F03-J0216-spec}} \hfill\resizebox{0.495\hsize}{!}{\includegraphics{F03-J1451-spec}} \\[-2ex]
  \resizebox{0.495\hsize}{!}{\includegraphics{F03-J0935-spec}} \hfill\resizebox{0.495\hsize}{!}{\includegraphics{F03-J1627-spec}} \\[-2ex]
  \resizebox{0.495\hsize}{!}{\includegraphics{F03-J0959-spec}} \hfill\resizebox{0.495\hsize}{!}{\includegraphics{F03-J2238-spec}} \\[-2ex]
  \resizebox{0.495\hsize}{!}{\includegraphics{F03-J1204-spec}} \hfill\resizebox{0.495\hsize}{!}{\includegraphics{F03-J2300-spec}} \\[-2ex]
  \resizebox{0.495\hsize}{!}{\includegraphics{F03-J1250A-spec}}\hfill\resizebox{0.495\hsize}{!}{\includegraphics{F03-J2303-spec}}
  \caption{Global spectra obtained by adding spectra within elliptical
    apertures with diameters between 5 and 14 spaxels, matched to the
    surface brightness of the objects. Wavelengths are in \AA. The
    three spectra top left are from the pilot programme and cover a
    different wavelength range then the spectra from the large
    programme. Emission lines from the sources are marked. The narrow
    residuals of sky lines at 5577\,\AA, 5890\,\AA\ and 6300\,\AA\
    have been interpolated over for display purposes.}
  \label{fig:spectra}
\end{figure*}

Fig.~\ref{fig:spectra} shows integrated spectra obtained by summing
the one-dimensional spectra from fibres within elliptical apertures
following the shape and size of each lens galaxy. Typical diameters of
these apertures are around $4$~arcsec, i.e.~a little larger than the
3~arcsec circular aperture used by the SDSS.

The spectra are typical for early-type galaxies, with strong metal
absorption lines, in particular \ion{Ca}{i} H~and K, the G and the
Mg\,b bands. With the exception of J2321 \citep{Czoske2008}, no
significant emission lines from the lens galaxies are detected. This
is true even for those galaxies that show clear disk and spiral
structure in the HST images.

Most of the spectra clearly show [\ion{O}{ii}]\,$\lambda 3727$
emission from the lensed source, although in some cases this line
falls outside the wavelength range of the VIMOS spectra. In some
cases, in particular J1250B and J1451, several Balmer emission lines
from the source can be seen. These become more prominent in the
residuals of the kinematic fits, described in
Sect.~\ref{sec:kinematic_analysis}.

\section{Kinematic analysis}
\label{sec:kinematic_analysis}

\subsection{Method}
\label{ssec:kinematic:method}

\begin{table}
  \centering
  \caption{Stellar templates used for kinematic analysis of the VIMOS/IFU sample}
  \begin{tabular}{lll}
    \hline\hline
    Galaxy & Template star & Spectral type \\
    \hline
    J\,0037  & HD\,249     & K1\,IV  \\
    J\,0216  & HD\,195506  & K2\,III \\
    J\,0912  & HD\,121146  & K3\,IV  \\
    J\,0935  & HD\,195506  & K2\,III \\
    J\,0959  & HD\,102328  & K3\,III \\
    J\,1204  & HD\,221148  & K3\,III \\
    J\,1250A & HD\,145328  & K1\,III \\
    J\,1250B & HD\,216640  & K1\,III \\
    J\,1251  & HD\,145328  & K1\,III \\
    J\,1330  & HD\,25975   & K1\,III \\
    J\,1443  & HD\,85503   & K2\,III \\
    J\,1451  & HD\,114092  & K4\,III \\
    J\,1627  & HD\,195506  & K2\,III \\
    J\,2238  & HD\,195506  & K2\,III \\
    J\,2300  & HD\,195506  & K2\,III \\
    J\,2303  & HD\,121146  & K3\,IV  \\
    J\,2321  & HD\,77818   & K1\,IV  \\
    \hline
  \end{tabular}
  \label{tab:templates:IndoUS}
\end{table}

Many methods to determine the line-of-sight velocity distribution
(LOSVD) of early-type galaxies from spectra have been proposed in the
past \citep[e.g.][]{Rix-White1992, vanderMarel-Franx1993,
  Cappellari-Emsellem2004}. We use the conceptually simplest method,
template-fitting in pixel space. Our method is implemented as a
package (\textsc{slacR}) in the statistical language \textsc{R}
\citep{R}\footnote{\texttt{http://www.r-project.org}. The
  \textsc{slacR} package can be obtained from the first author on
  request.}, following descriptions from several authors, in
particular \citet{vanderMarel-Franx1993} and
\citet{Kelson2000}. Compared to the version used in
\citet{Czoske2008}, we have modified and extended our method
sufficiently to warrant a full description of the method here.

The model parameters $\bmath \theta$ are determined by minimizing the merit
function
\begin{equation}
  \label{eq:merit_function}
  S = \sum_{i} \frac{1}{\sigma_{i}^{2}} \left(s_{i} - \hat{s}(\lambda_{i},
    {\bmath \theta})\right)^{2}\,,
\end{equation}
where the sum extends over all ``good'' pixels $\lambda_{i}$ in the
observed spectrum $s_{i}$. The observed spectrum is transformed to the
rest frame of the respective lens galaxy prior to the analysis by
dividing observed wavelengths by a factor $1+z_{\mathrm{lens}}$. The
effective resolution of the spectrum is increased to
$R(1+z_{\mathrm{lens}})$ compared to the nominal resolution $R$ of the
spectrograph (Sect.~\ref{ssec:obs:VIMOS}).

The data-model $\hat{s}$ (in vector-matrix notation) is given by the
equation
\begin{equation}
  \label{eq:kinematic_model}
  \hat{s}_{i} =  [ \bmath G *\bmath t ](\lambda_{i})\, p^{(m)}(\lambda_{i}) +
  q^{(n)}(\lambda_{i}) + \epsilon_{i}\,.
\end{equation}
Here, the vector $ \bmath G$ is the LOSVD, taken to be a Gaussian in
velocity, with kinematic parameters $v$ (streaming motion) and
$\sigma_{v}$ (velocity dispersion). The vector $\bmath t$ is a stellar
template spectrum, taken from the Indo-US library of stellar spectra
\citep{Valdes2004}. Templates were chosen by fitting a random sample
of IndoUS spectra to the aperture-integrated VIMOS/IFU spectra and
selecting one of the best-fitting (in the least-squares sense)
template candidates. We did not always choose the template giving the
absolute minimum $\chi^{2}$: It was noticed that HD\,195506 appeared
among the best-fitting templates for several systems and was therefore
chosen for all of these. Table~\ref{tab:templates:IndoUS} lists the
templates used for the kinematic fits, as well as their spectral type
and luminosity class. As expected, the selected template stars are
predominantly late-type giants. The same template was used for fitting
all the individual fibre spectra for each system. The effect of
template mismatch on the derived kinematic parameters will be
discussed below.

\begin{figure}
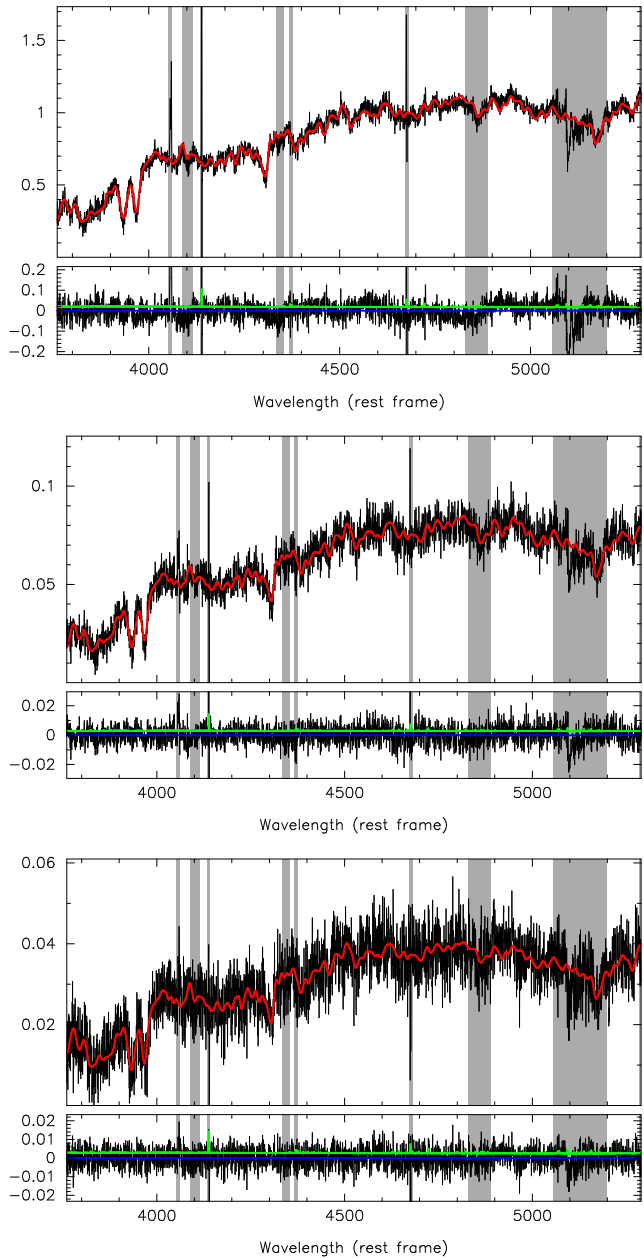

  \centering
  \resizebox{\hsize}{!}{\includegraphics{F04-J0935_global_fit}}\\
  \vspace*{2ex}
  \resizebox{\hsize}{!}{\includegraphics{F04-J0935-21-25-SN23}}\\
  \vspace*{2ex}
  \resizebox{\hsize}{!}{\includegraphics{F04-J0935-20-26-SN12}}
  \caption{Kinematic fitting of spectra of system J0935. Top: global
    spectrum summed in an aperture of radius 2.3\,arcsec, $S/N \approx
    45$ per wavelength bin. Centre: single spaxel with $S/N\approx 23$
    per bin. Bottom: single spaxel with $S/N\approx 12$ per bin. The
    top panel of each pair shows the spectrum and the template fit in
    red. The bottom panel shows the residuals with the expected noise
    level in green (Appendix~\ref{sec:noise_estimation}). The grey
    areas mark regions that were excluded from the fits. Wavelengths
    are given in the rest frame of the lens ($z=0.3475$). The
    [\ion{O}{ii}] line from the source can be seen at 4057\,\AA\ in
    this plots.}
  \label{fig:kinematic_fits}
\end{figure}

For the kinematic fit, the template is first brought to the effective
instrumental resolution of the observed spectrum by convolving with a
kernel that is a Gaussian in wavelength. The convolution with the
LOSVD $\bmath G$ is performed in $\log \lambda$ (equivalently:
velocity), and the result is then resampled to the same wavelength
grid as the data $\bmath s$, thus avoiding having to resample the data
beyond what has been done during the data reduction.  The functions
$p^{(m)}(\lambda)$ and $q^{(n)}(\lambda)$ are multiplicative and
additive correction polynomials of order $m$ and $n$,
respectively. These polynomials are needed to correct any
low-frequency differences between the galaxy and template spectra due
to insufficient flux calibration, contamination by the continuum of
the lensed background galaxy and other effects which are not related
to the kinematics of the lens galaxy. In contrast to
\citet{Czoske2008}, where only a short wavelength range around a
single spectral feature was used and a linear correction function was
sufficient, we here use the full wavelength range. The coefficients
are determined by a linear fit nested within the non-linear
optimization for the kinematic parameters $v$ and $\sigma_{v}$. The
orders of the polynomials are determined by inspecting large-scale
systematics in the residuals from the fits; we obtain satisfactory
corrections for $m=4$ and $n=6$.

A number of features in the spectra were excluded in the kinematic
fits. Possibly present Balmer lines in the lens spectrum are due to
younger stellar populations that are not adequately described by the
late-type template stars. The strength of the \ion{Mg}{b} band and
Na\,D lines is enhanced compared to that in the galactic stars used as
templates \citep{Barth2002}. Also masked were night-sky emission lines
and atmospheric absorption features, as well as emission lines from
the lensed background sources. Finally, a 3-$\sigma$ clipping
algorithm (three iterations) was applied to detect possible outliers.

The noise $\bmath \epsilon$ is assumed to be normally distributed with
mean zero and wavelength-dependent dispersion $\bmath
\sigma_{\epsilon}$.  Since \textsc{vipgi} does not produce noise
estimates that take the details of the reduction procedure into
account, we have to estimate the noise spectra after the fact, using a
model including readout noise and photon noise as described in
Sect.~\ref{sec:data_reduction} and
Appendix~\ref{sec:noise_estimation}. The average signal-to-noise ratio
used to characterize a spectrum was determined by dividing the
spectrum by its noise spectrum and taking the median. Signal-to-noise
ratios are thus given per wavelength pixel of width 0.644\,\AA.

The noise was propagated to error estimates on the kinematic
parameters using a Monte-Carlo method. Gaussian noise was added to the
best-fit model with wavelength-dependent dispersion given by the
corresponding noise spectrum and the resulting spectrum was fitted in
the same way as the original spectrum. The quoted errors on the
parameters of the original spectrum are the 16\% and 84\% percentiles
of the kinematic parameters $v$ and $\sigma_{v}$ obtained from 300
such realizations.

The errors due to template mismatch can be estimated from the
distribution of best-fit values determined with 755 stellar spectra of
widely varying stellar type. Reasonable candidate templates are
defined for this purpose as those giving $\chi^{2} <
\chi^{2}_{\mathrm{min}} + 1$, where $\chi^{2}_{\mathrm{min}}$ is the
minimum value obtained with the set of stellar spectra. The rms values
of $\sigma_{v}$ obtained with these template candidates lie between
5~and 10\,\% (maximum value is 10.6\,\% for J0959). For the
two-dimensional kinematic maps, template mismatch will mostly have an
effect on the overall level, but not on the structure if stellar
populations across the galaxy are homogeneous.

\subsection{Results}
\label{ssec:kinematic:results}

Example fits for an aperture-integrated spectrum and spectra from
individual fibres are shown in Fig.~\ref{fig:kinematic_fits}. The
template fits generally reproduce the observed spectra to the expected
noise level.

\begin{figure*}
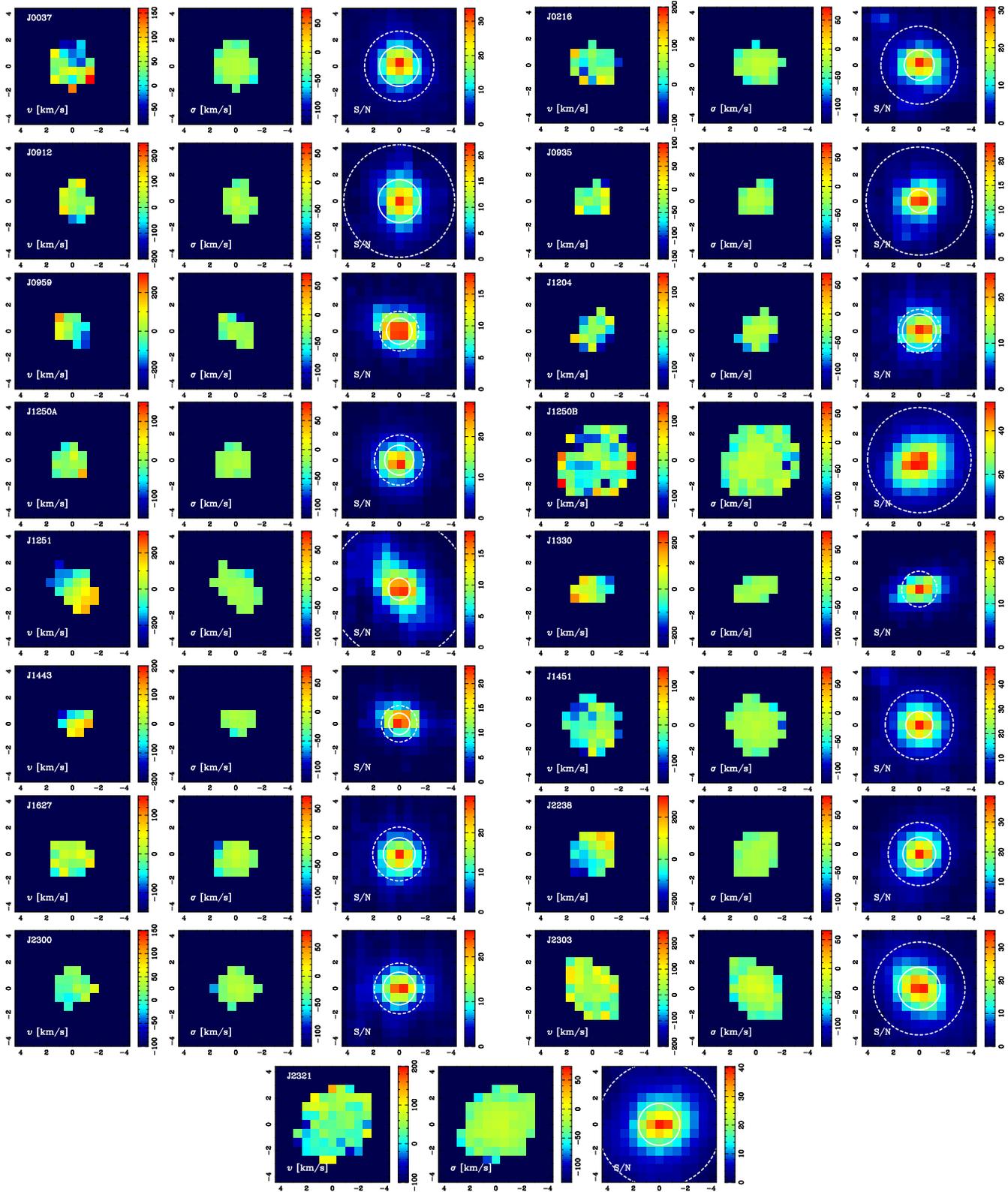

  \centering
  \resizebox{0.48\hsize}{!}{\includegraphics{F05-J0037-kinmap}}\hfill  \resizebox{0.48\hsize}{!}{\includegraphics{F05-J0216-kinmap}}
  \resizebox{0.48\hsize}{!}{\includegraphics{F05-J0912-kinmap}}\hfill  \resizebox{0.48\hsize}{!}{\includegraphics{F05-J0935-kinmap}}
  \resizebox{0.48\hsize}{!}{\includegraphics{F05-J0959-kinmap}}\hfill  \resizebox{0.48\hsize}{!}{\includegraphics{F05-J1204-kinmap}}
  \resizebox{0.48\hsize}{!}{\includegraphics{F05-J1250A-kinmap}}\hfill \resizebox{0.48\hsize}{!}{\includegraphics{F05-J1250B-kinmap}}
  \resizebox{0.48\hsize}{!}{\includegraphics{F05-J1251-kinmap}}\hfill  \resizebox{0.48\hsize}{!}{\includegraphics{F05-J1330-kinmap}}
  \resizebox{0.48\hsize}{!}{\includegraphics{F05-J1443-kinmap}}\hfill  \resizebox{0.48\hsize}{!}{\includegraphics{F05-J1451-kinmap}}
  \resizebox{0.48\hsize}{!}{\includegraphics{F05-J1627-kinmap}}\hfill  \resizebox{0.48\hsize}{!}{\includegraphics{F05-J2238-kinmap}}
  \resizebox{0.48\hsize}{!}{\includegraphics{F05-J2300-kinmap}}\hfill  \resizebox{0.48\hsize}{!}{\includegraphics{F05-J2303-kinmap}}
  \resizebox{0.48\hsize}{!}{\includegraphics{F05-J2321-kinmap}}
  \caption{Results of kinematic analysis of the SLACS/IFU sample. For
    each target, the left plot shows the map of systematic velocity
    in the lens galaxy and the middle plot shows the map
    of velocity dispersion. All velocities refer to the rest
    frame of the lens galaxy. The right plots show the mean
    signal-to-noise ration (per pixel of width $0.65\,$\AA) of the
    spectra. The axis labels are in arcsec. The maps are oriented such
    that north is to the top and east to the left.}
  \label{fig:Kinematic_maps}
\end{figure*}

Fig.~\ref{fig:Kinematic_maps} shows the maps of velocity, velocity
dispersion and spectral signal-to-noise. The kinematic maps are
restricted to spectra with an average signal-to-noise ratio $S/N > 8$
(per pixel) for which kinematic parameters can be reliably
measured. Two-thirds of the sample show little structure in the maps
and display kinematics typical for nearly-isothermal
pressure-supported slow rotators. Clear rotation patterns can be
discerned in the velocity maps for the remaining third of the sample
(e.g.~J0959, J1251 or J2238). Kinematic maps of local galaxies in
general show a central peak of velocity dispersion; at the redshifts
of our sample, the spatial resolution of the integral-field
spectrograph is not sufficient to resolve this peak.

\begin{figure*}
  \centering
  \includegraphics{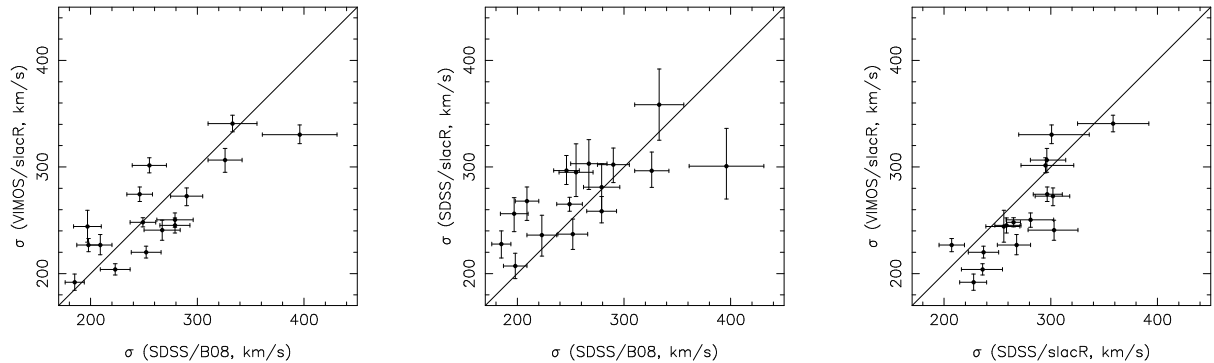}
  \caption{Comparison of velocity dispersion measurements. The VIMOS
    spectra used here are summed over all spaxels within the same
    aperture as the SDSS spectra and analysed using our method
    (\textsc{slacR}). For the SDSS spectra, values measured with
    \textsc{slacR} and values obtained from \citet[B08]{Bolton2008a} are
    used.}
  \label{fig:sigma_comparison}
\end{figure*}

Fig.~\ref{fig:sigma_comparison} compares our velocity dispersion
measurements on the global (aperture-integrated) VIMOS spectra to
measurements on the spectra from the Sloan Digital Sky Survey
(Table~\ref{tab:sample}). The agreement between the different
measurements is generally good, which gives us confidence in the
quality of the spectra and the reliability of the analysis method.

For the SDSS spectra we compare two independent measurements: the
values listed by \citet{Bolton2008a} and new values determined with
our code using the same wavelength ranges and template spectra as for
the VIMOS spectra. The middle panel of Fig.~\ref{fig:sigma_comparison}
compares these measurements, which differ only in the analysis method.
The error bars on the slacR measurements include only the effect of
noise in the spectra and do not take into account systematic effects
such as choice of template and other parameters. The good agreement
within the errors shows that systematic effects contribute only little
to the uncertainties. A notable outlier is J0935 for which we measure
$\sigma = 330\,\mathrm{km\,s^{-1}}$ compared to
$396\,\mathrm{km\,s^{-1}}$ from \citet{Bolton2008a}.

The right panel compares measurements on the SDSS spectra to those on
the VIMOS spectra, using identical methods. The different instrumental
resolution ($R=1800$ for SDSS, $R=2500$ for VIMOS) has been taken into
account. There appears to be a slight offset of the velocity
dispersions obtained from the SDSS spectra compared to those from the
VIMOS spectra. A similar trend is visible in the middle panel, while
the comparison of the VIMOS measurements to the SDSS measurements from
\citet{Bolton2008a} does not show any offset.

\section{Secondary sources}
\label{sec:other_sources}

Some of the VIMOS/IFU fields in this sample contain objects in
addition to the primary targets. For these fields we show larger-scale
cut-outs from SDSS $r$-band images and label them by number. We
extracted aperture spectra from the VIMOS data cubes and determined
their redshifts by cross-correlation \citep{Tonry-Davis1979} with a
set of SDSS template spectra. Coordinates relative to the primary
target and redshifts are listed in Table~\ref{tab:other_sources}. The
table also lists redshifts for the lens galaxies obtained by the same
method in order to ensure that relative redshifts are accurate. Most
of the additional objects are consistent with being at the same
redshift as the lens galaxy. 

\begin{table}
  \centering
  \caption{Data for additional sources in VIMOS/IFU
    fields. Coordinates are in arcsec relative to the position of the
    lens galaxy (always labelled as ``1'') in each field. Redshifts
    marked with ``?'' are suggested by peaks in the cross-correlation
    but cannot be confirmed unambiguously by visual
    inspection. Objects with $z=0$ are stars.}
  \begin{tabular}{lrrl}
    \hline\hline
    Object & \multicolumn{1}{c}{$\Delta \alpha\,(\arcsec)$} & \multicolumn{1}{c}{$\Delta \delta\,(\arcsec)$} & \multicolumn{1}{c}{z} \\
    \hline
    J0216-1 & $  0  $ & $ 0  $ & 0.3311 \\
    J0216-2 & $ +2.9$ & $+3.2$ & 0.3327 \\
    J0216-3 & $ -0.9$ & $-9.2$ & 0.3339 \\
    J0216-4 & $-13.1$ & $+3.5$ & 0.3327 \\
    J0216-5 & $+12.4$ & $+4.8$ & 0      \\
    \hline
    J0912-1 & 0 & 0 & 0.1638 \\
    J0912-2 & $+4.3$ & $+6.4$ & 0.1609 \\
    \hline
    J0935-1 & $ 0  $ & $ 0  $ & 0.3468 \\
    J0935-2 & $+0.6$ & $+5.2$ & 0.3407 \\
    J0935-3 & $+8.6$ & $-5.3$ & 0.3528 \\
    \hline
    J1250B-1 & $ 0  $ & $  0  $ & 0.0866 \\
    J1250B-2 & $-5.1$ & $+11.1$ & 0.2621? \\
    \hline
    J1251-1 & $  0  $ & $  0  $ & 0.2238 \\
    J1251-2 & $+15.2$ & $+10.8$ & 0 \\
    \hline
    J1451-1 & $  0  $ & $ 0  $ & 0.1248 \\
    J1451-2 & $-10.5$ & $-4.8$ & 0 \\
    J1451-3 & $ +2.9$ & $+3.7$ & 0.5196? \\
    \hline
    J2238-1 & $ 0  $ & $ 0  $ & 0.1365 \\
    J2238-2 & $+4.4$ & $-4.6$ & 0.1358 \\
    \hline
    J2300-1 & $  0  $ & $ 0  $ & 0.2280 \\
    J2300-2 & $-11.5$ & $-7.1$ & 0.2277? \\
    \hline
  \end{tabular}
  \label{tab:other_sources}
\end{table}

\begin{figure*}
  \centering
  \resizebox{\hsize}{!}{
    \includegraphics{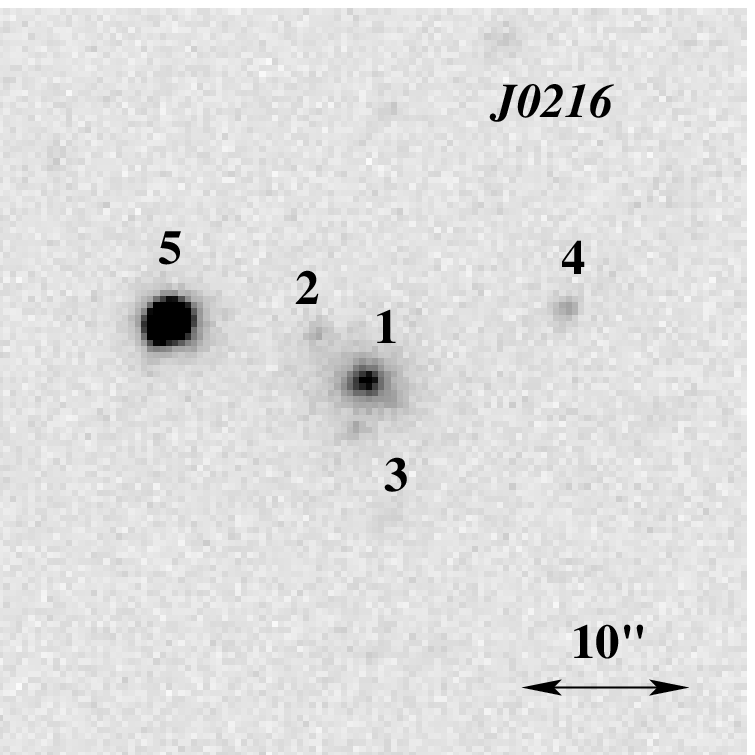}
    \includegraphics{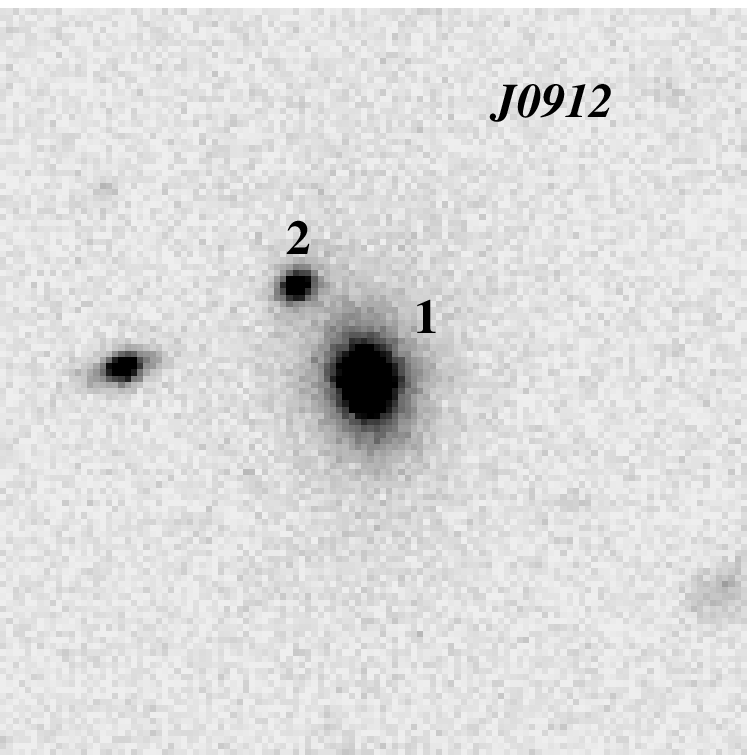}
    \includegraphics{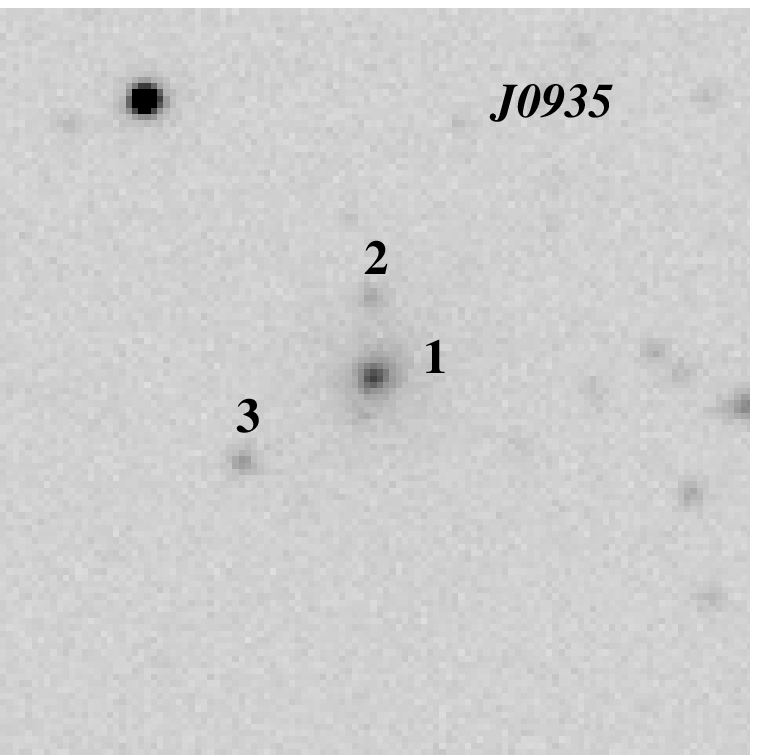}
    \includegraphics{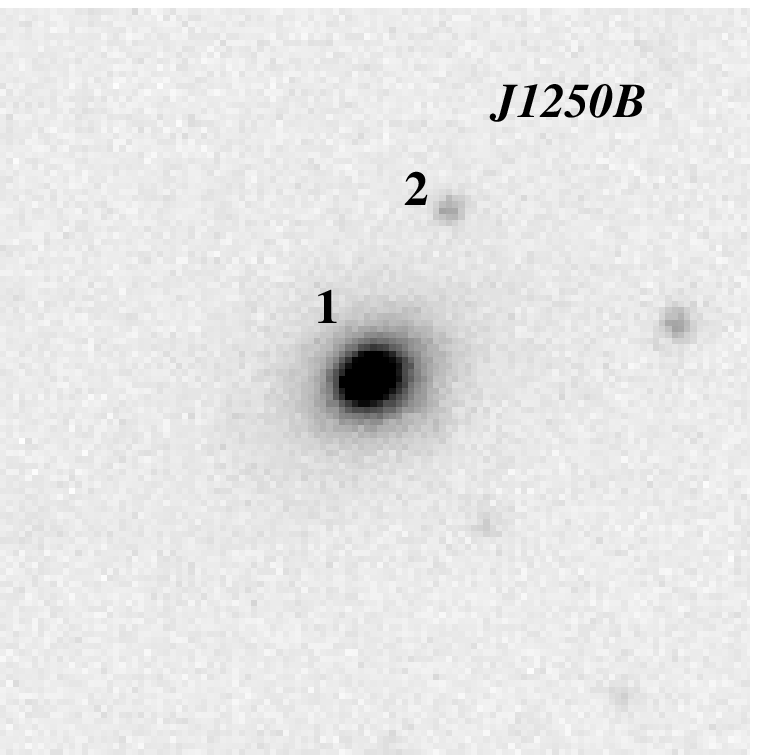}
  }
  \resizebox{\hsize}{!}{
    \includegraphics{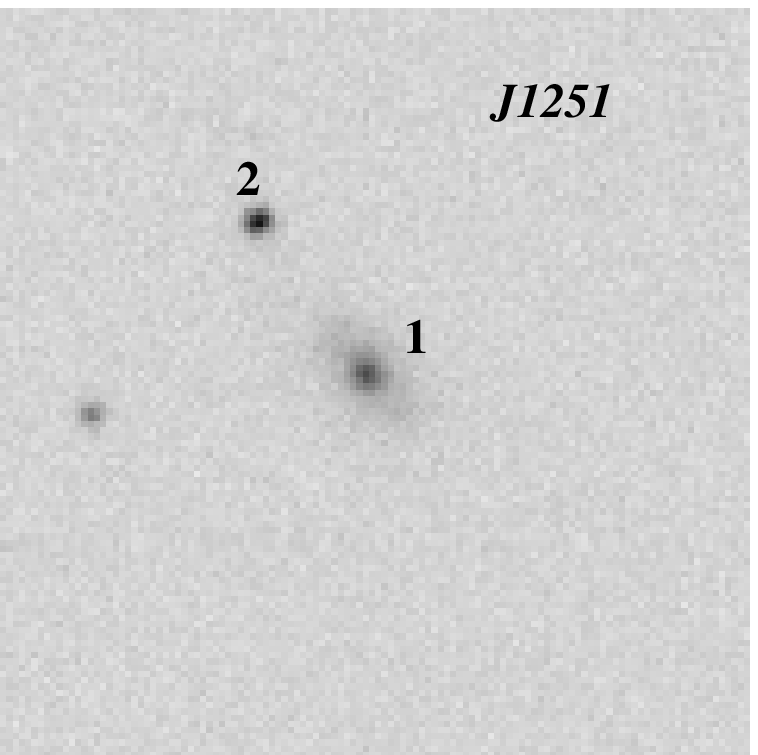}
    \includegraphics{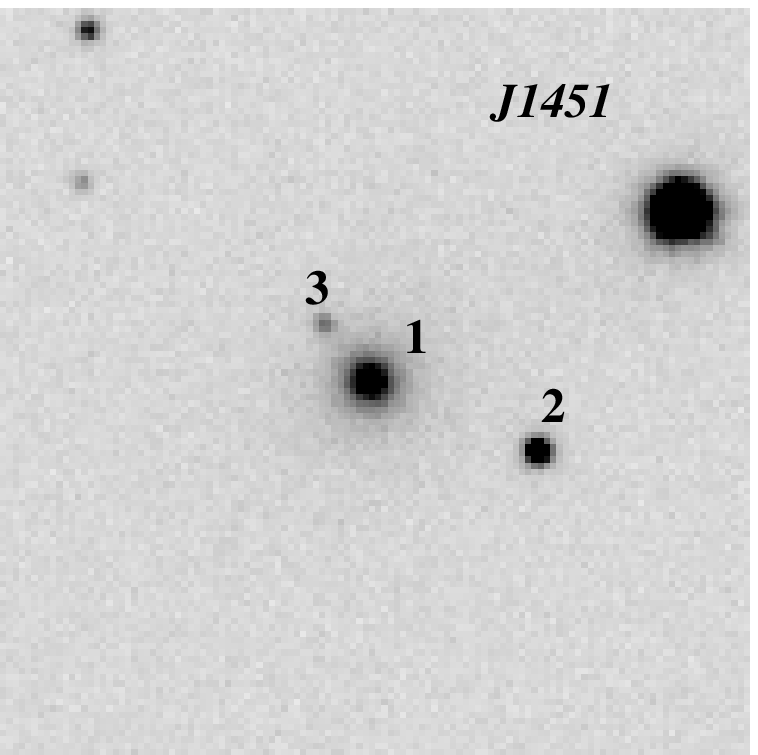}
    \includegraphics{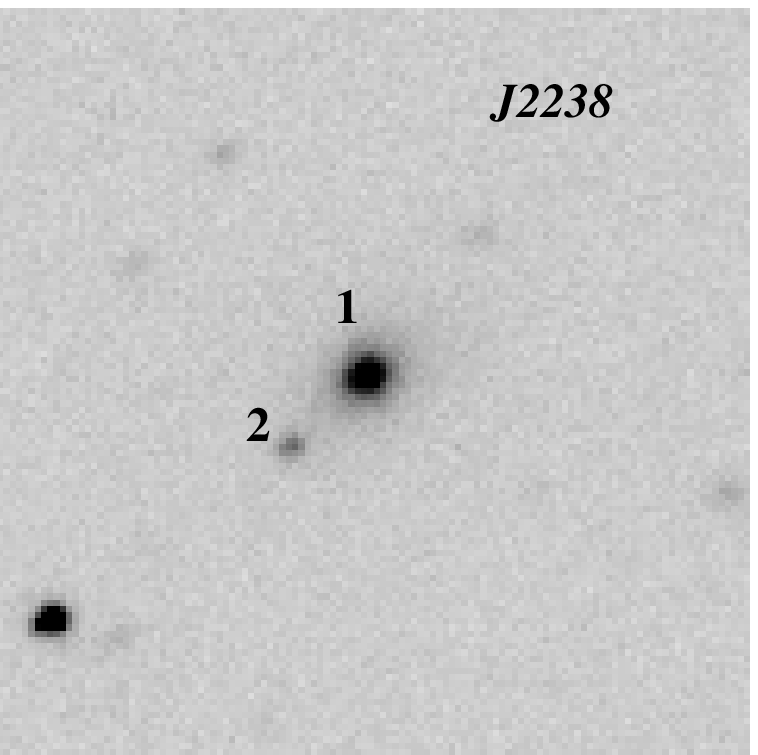}
    \includegraphics{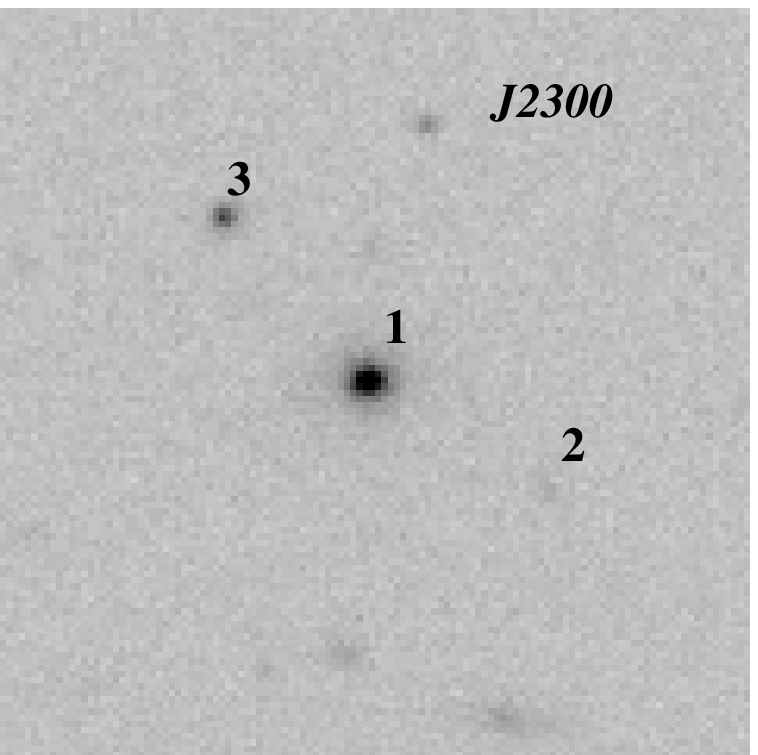}
  }
  \caption{Finding charts for the sources listed in
    Table~\ref{tab:other_sources}. Images are $r$-band images from the
    SDSS}
  \label{fig:other_sources:charts}
\end{figure*}

\section{Conclusions}
\label{sec:conclusions}

In this paper, we have presented integral-field spectroscopic data
obtained with VIMOS/IFU on a sample of 17~early-type lens galaxies
selected from the SLACS survey. The data permit spatially resolved
reconstruction of the stellar kinematics in these galaxies, presented
as maps of systematic velocity and velocity dispersion. The sample,
which spans a redshift range of $0.08$ to $0.35$ for the lens
galaxies, is well-suited for a comparison with local samples, such as
SAURON \citep{Emsellem2004} and ATLAS\textsuperscript{3D}
\citep{Cappellari2010short}.

The distinction of slow-rotating and fast-rotating galaxies found in
the SAURON sample \citep{Emsellem2007} and earlier work
\citep[e.g.][]{Davies1983} is also evident from the kinematic maps for
our sample. About a third of the galaxies in the sample show clear
evidence for rotation in their velocity maps. 

While the spatial resolution of integral-field spectroscopic
observations of galaxies at cosmological distances is necessarily much
coarser than for local galaxies, the present sample offers the unique
advantage that all the galaxies act as gravitational lenses and
therefore offer the possibility to model their mass distribution using
the two complementary methods of stellar dynamics and gravitational
lensing. A Bayesian implementation of a fully self-consistent
modelling algorithm for the joint analysis of stellar dynamics and
gravitational lensing was developed by \citep{Barnabe-Koopmans2007}
and applied to subsets of this sample by \citet{Czoske2008} and
\citet{Barnabe2009b}. Modelling and analysis of the full data set is
presented in \citet{Barnabe2011}.

\section*{Acknowledgments}

The data published in this paper have been reduced using
\textsc{vipgi}, designed by the VIRMOS Consortium and developed by
INAF Milano. M.B. acknowledges the support from an NWO programme
subsidy (project number 614.000.417) from the Department of Energy
contract DE-AC02-76SF00515. O.C. and L.V.E.K. were supported (in part)
through an NWO-VIDI programme subsidy (project number
639.042.505). T.T. acknowledges support from the NSF through CAREER
award NSF-0642621 and by the Packard Foundation through a Packard
Fellowship.

\bibliography{Czoske2011-SLACS}

\appendix

\section{Estimation of noise vectors}
\label{sec:noise_estimation}

Noise vectors can be estimated by following the reduction process from
the initial raw data through the final data product. Using a model
that takes into account photon noise (from object and sky) and readout
noise, it is possible to express the variance in the data at each step
in the reduction as a simple rescaling of the current intermediate
data product. 

The raw image frames have values $\vec{d}$, expressed in ADU; these
contain in addition to the useful data the bias $\vec{b}$. The noise
variance is the quadratic sum of the photon noise due to the data
values $\vec{d} - \vec{b}$ and the readout noise $r$, given in
electrons. In terms of \emph{electrons}, the signal (without bias) is
$(\vec{d} - \vec{b})/g$ and the variance is thus
\begin{equation}
  \label{eq:noise_1}
  \widehat{\sigma}_{\mathrm{raw}}^{2} = \frac{1}{g}\,(\vec{d} - \vec{b}) + r^{2}\quad[\mathrm{e}^{-}].
\end{equation}
The conversion from electrons to ADU is here given by the gain factor
$g$ (in ADU/e\textsuperscript{$-$}).

Variance scale as
\begin{equation}
  \label{eq:variance_rescale}
  \sigma_{2}^{2} = c^{2} \sigma_{1}^{2}
\end{equation}
when the data are rescaled according to $f_{2} = c f_{1}$. These
relations preserve the signal-to-noise under rescaling. The variance
of the raw images in terms of ADU is thus
\begin{equation}
  \label{eq:noise_2}
  \sigma_{\mathrm{raw}}^{2} = g\,(\vec{d} - \vec{b}) + R^{2}\quad \mathrm{[ADU]},
\end{equation}
where $R = gr$ is the readout noise in ADU. 

The preliminary reduction step in \texttt{vipgi} subtracts the bias
frame and produces \texttt{BFC} files. Compared to the raw frame, the
variance is increased by the noise in the combined bias, $\sim
R^{2}/5$. Since the readout noise is small anyway, we neglect this
term.
\begin{eqnarray}
  \label{eq:value_BFC}
  \mathrm{BFC} & = & \vec{d} - \vec{b}\\
  \label{eq:noise_BFC}
  \sigma_{\mathrm{BFC}}^{2} & = & g\,\mathrm{BFC} + R^{2}\,. 
\end{eqnarray}

The following reduction step applies the wavelength calibration,
rectifies the two-dimensional spectra, and extracts one-dimensional
spectra from these. The output files are names \texttt{BFCEx}. The
\texttt{EXR2D} extension contains the rectified version of
\texttt{BFC}. Neglecting the correlation of pixel noise introduced by
the resampling, we have, as for the \texttt{BFC}:
\begin{equation}
  \label{eq:noise_EXR2D}
  \sigma_{\mathrm{EXR2D}}^{2} = g\,\mathrm{EXR2D} + R^{2}\,.
\end{equation}

The 1D spectra in \texttt{EXR1D} are created from \texttt{EXR2D} by
the Horne optimal extraction algorithm
\citep{Horne1986,Zanichelli2005short}, which sums over the spatial
direction of the 2D spectrum using an estimate of the spatial profile
of the light distribution as weights. The algorithm permits direct
estimation of the noise variance in the extracted spectra; this is,
however, not implemented in \texttt{vipgi}. We replace the Horne
variance by the variance of a straight sum over $N_{c}=5$ columns in
the 2D spectrum:
\begin{eqnarray}
  \label{eq:value_EXR1D}
  \mathrm{EXR1D} & = & \sum_{N_{\mathrm{c}}} \mathrm{EXR2D} \\
  \label{eq:noise_EXR1D}
  \sigma_{\mathrm{EXR1D}}^{2} & = & \sum_{N_{\mathrm{c}}}
  \sigma_{\mathrm{EXR2D}}^{2} 
   =  \sum_{N_{\mathrm{c}}} (g\,\mathrm{EXR2D} +  R^{2}) \nonumber\\
  & = & g\,\mathrm{EXR1D} + N_{c} R^{2}
\end{eqnarray}
Comparison of the noise predicted by Eq.~(\ref{eq:noise_EXR1D}) to
real data in pure sky fibres (no object contribution) indicates that
the noise is indeed somewhat overestimated, although not dramatically
so.

The relative fibre transmissivity $t$ of a given fibre is estimated
from the measured flux in a sky emission line and stored in a FITS
table. Its application results in
\begin{eqnarray}
  \label{eq:value_EXR1DTC}
  \mathrm{EXR1DTC} & = &\frac{1}{t}\,\mathrm{EXR1D}\\
  \label{eq:noise_EXR1DTC}
  \sigma_{\mathrm{EXR1DTC}}^{2} & = &
  \frac{1}{t^{2}}\,\sigma_{\mathrm{EXR1D}}^{2}\\ 
  & = &  \frac{g}{t}\,\frac{1}{t}\,\mathrm{EXR1D} +
  \frac{N_{\mathrm{c}}}{t^{2}}\,R^{2}\\
  & = & \frac{g}{t}\,\mathrm{EXR1DTC} +
  \frac{N_{\mathrm{c}}}{t^{2}}\,R^{2}\,.
\end{eqnarray}
The flux calibration uses the sensitivity function $s(\lambda)$. It
acts like any other rescaling. The sensitivity function is normalized
to an exposure time of $1\,\mathrm{s}$, hence it has to be rescaled to
exposure time $T$ (in seconds). The estimate of the sky is based on a
large number of fibers and has negligible effect on the variance.
\begin{eqnarray}
  \label{eq:value_EXR1DFLUX} 
  \mathrm{EXR1DFLUX} & = & \frac{1}{Ts}\left(\mathrm{EXR1DTC} -
    \mathrm{SKY1D}\right) \\
  \label{eq:noise_EXR1DFLUX}
  \sigma_{\mathrm{EXR1DFLUX}}^{2} & = &
  \frac{g}{Tst}\,\frac{1}{Ts}\,\mathrm{EXR1DTC} +
  \frac{N_{\mathrm{c}}}{T^{2}s^{2}t^{2}}\,R^{2}\\
  & = &\frac{g}{Tst}\,\overline{\mathrm{EXR1DFLUX}} +
  \frac{N_{c}}{T^{2} s^{2} t^{2}}\,R^{2} \,.
\end{eqnarray}
The overline indicates that the sky has not been subtracted.

The final data cube is generated by combining $N_{\mathrm{exp}}$
exposures using the median which results in a somewhat larger variance
than the arithmetic mean. Simplifying again, we use the variance for
the latter:
\begin{eqnarray}
  \label{eq:value_cube}
  \mathrm{CUBE} & \approx & \frac{1}{N_{\mathrm{exp}}}\,\sum_{N_{\mathrm{exp}}}
  \mathrm{EXR1DFLUX} \\
  \label{eq:noise_cube}
  \sigma_{\mathrm{CUBE}}^{2} &\approx &
  \frac{1}{N_{\mathrm{exp}}^{2}}\,\sum_{N_{\mathrm{exp}}}
  \sigma_{\mathrm{EXR1DFLUX}}^{2} \\
  & = & \frac{1}{N_{\mathrm{exp}}^{2}}\,
  \sum_{\mathrm{N_{\mathrm{exp}}}}\frac{g}{Tst}\,
  \overline{\mathrm{EXR1DFLUX}} +
  \frac{N_{\mathrm{c}}}{N_{\mathrm{exp}}}\,\frac{R^{2}}{T^{2}s^{2}t^{2}}\\
  & = &
  \frac{1}{N_{\mathrm{exp}}}\,\frac{g}{Tst}\,\overline{\mathrm{CUBE}} +
  \frac{N_{\mathrm{c}}}{N_{\mathrm{exp}}}\,\frac{R^{2}}{T^{2}s^{2}t^{2}}
\end{eqnarray}
Note that due to the dithering between exposures some of the spaxels
in the cube arise from a combination of fibres from different
quadrants which have different gain $g$ and sensitivity functions
$s$. Assuming that we can use some effective values averaged over the
four quadrants, the factor $g/st$ in the second row of
Equation~(\ref{eq:noise_cube}) can be taken out of the sum. This way,
the variance of the cube can be obtained from a mock cube
$\overline{\mathrm{CUBE}}$ which is produced in the same manner as the
actual cube $\mathrm{CUBE}$, only with sky subtraction turned off.

All the parameters in Eq.~(\ref{eq:noise_cube}) are stored as FITS
header keywords or in table extensions of \texttt{vipgi} products. 

In the final step, the global spectrum is created by summing over
$N_{\mathrm{fib}}$ spaxels (defined by an aperture) in the final data
cube, giving
\begin{eqnarray}
  \label{eq:value_SPEC}
  \mathrm{SPEC} & = & \sum_{N_{\mathrm{fib}}}\,\mathrm{CUBE}\\
  \label{eq:noise_SPEC}
  \sigma_{\mathrm{SPEC}}^{2} & = & \sum_{N_{\mathrm{fib}}}
  \sigma_{\mathrm{CUBE}}^{2} \\
  & = &
  \frac{1}{N_{\mathrm{exp}}}\,\frac{g}{Tst}\,\sum_{N_{\mathrm{fib}}}
  \overline{\mathrm{CUBE}} + \frac{N_{\mathrm{fib}}
    N_{\mathrm{c}}}{N_{\mathrm{exp}}}\,\frac{g^{2}r^{2}}{T^{2}s^{2}t^{2}}
  \\
  & = &
  \frac{1}{N_{\mathrm{exp}}}\,\frac{g}{Tst}\,\overline{\mathrm{SPEC}}
  + \frac{N_{\mathrm{fib}}
    N_{\mathrm{c}}}{N_{\mathrm{exp}}}\,\frac{g^{2}r^{2}}{T^{2}s^{2}t^{2}}
\end{eqnarray}

Equations~(\ref{eq:value_SPEC}) and~(\ref{eq:noise_SPEC}) provide the
recipes for the final spectrum. The readout noise $r$ is written in
terms of electrons, as it usually is given in the fits headers.

\label{lastpage}

\clearpage

\end{document}